\documentclass{pazhb_eng} 
     
\usepackage{graphicx} 
\usepackage{amsmath} 
\usepackage{epsfig,graphics} 
\usepackage{rotating} 
\usepackage{ulem}

\usepackage{epstopdf}
\usepackage{pdfpages}

\usepackage{longtable}
\usepackage{caption}
\usepackage{dcolumn}
\usepackage{pdflscape}

\usepackage{hyperref}

\usepackage{comment}
\usepackage{ulem}
\usepackage{color}

\newcommand {\be}{\begin{equation}} 
\newcommand {\ee}{\end{equation}}

\def\red#1{\textcolor{red}{#1}}
\def\БР#1{\red{\textbf{#1}}}

\newcommand{\zspec}{z_{\mbox{\scriptsize spec}}}
\newcommand{\zphot}{z_{\mbox{\scriptsize phot}}}

\newcommand\blfootnote[1]{
	\begingroup
	\renewcommand\thefootnote{}\footnote{#1}
	\addtocounter{footnote}{-1}
	\endgroup
	}

\def\2MASS{\textit{2MASS}}
\def\WISE{\textit{WISE}}
\def\3XMMDR4{\textit{3XMM-DR4}}
\def\SDSS{\textit{SDSS}}

\def\EAZY{\textit{EAZY}}
\def\Zphzsp{$|\zphot - \zspec |/(1+\zspec )$ }

\def\XMM{\textit{XMM-Newton}}

\def\K16{\textit{K16}}

\newcommand*{\PathtoSpec}{IMAGES}
\newcommand*{\PathtoSpecImage}{IMAGES}

\newcommand{\Voutl}{9} 

\newcommand{\Nobs}{18}
\newcommand{\Azt}{AZT-33IK}

\begin{document} 

\journalinfo{2017}{43}{3}{135}[145]
\UDK{524.7}

\title{Optical Spectroscopy of Candidates for Quasars at~3<\mbox{\small z}<5.5 from the XMM-Newton X-ray Survey.
A~Distant~X-ray~Quasar~at~\mbox{\small z}=5.08}

\author{\bf 
  G.A.~Khorunzhev\email{horge@iki.rssi.ru}\address{1}, R.A.~Burenin\address{1}, S.Yu.~Sazonov\address{1}, A.L.~Amvrosov\address{2}, M.V.~Eselevich\address{2}   
  \addresstext{1}{Space Research Institute, Russian Academy of Sciences, Profsoyuznaya ul. 84/32, Moscow, 117997 Russia}
  \addresstext{2}{Institute of Solar–Terrestrial Physics, Russian Academy of Sciences, Siberian Branch,\\
P.O.~Box 4026, Irkutsk, 664033 Russia} 
  }
  
\shortauthor{Khorunzhev~et~al.}  

\submitted{9.09.2016 г.}
   
\begin{abstract}  
We present the results of optical spectroscopy for 19 quasar candidates at photometric redshifts
$\zphot \gtrsim 3$, \Nobs \  of which enter into the Khorunzhev et al.~(2016) catalog (K16).
This is a catalog of quasar candidates and known type 1 quasars selected among the X-ray sources of the \textit{3XMM-DR4}catalog of the XMM-Newton serendipitous survey. We have performed spectroscopy for a quasi-random sample of new
candidates at the 1.6-m \Azt \  telescope of the Sayan Solar Observatory and the 6-m BTA telescope
of the Special Astrophysical Observatory. The spectra at \Azt \ were taken with the new low-
and medium-resolution ADAM spectrograph that was produced and installed on the telescope in 2015.
Fourteen of the \Nobs \ candidates actually have turned out to be quasars; 10 of them are at spectroscopic
redshifts $\zspec >3$. The high purity of the sample of new candidates suggests that the purity of the entire
K16 catalog of quasars is probably 70--80\%. One of the most distant ($\zspec=5.08$) optically bright
($i^\prime\lesssim 21$) quasars ever detected in X-ray surveys has been discovered. 

\end{abstract}

\section{1. Introduction}

Searching for quasars at $z\gtrsim 3$ is one of the most
important elements of studying the growth history of supermassive black holes and the evolution of massive 
galaxies in the Universe.
To construct the X-ray luminosity function for quasars at $z\gtrsim 3$ requires
collecting a large and well-defined X-ray sample of such objects with fluxes $\lesssim10^{-14}$~erg~s$^{-1}$cm$^{-2}$ (0.5--2~keV). 

The number of sources in the deep \XMM \ and \textit{Chandra} X-ray surveys 
(typical fluxes $\lesssim10^{-15}$ erg/s/cm$^2$ in the 0.5--2 keV energy band and areas of $\sim$1~sq.~deg.)
turns out to be insufficient to trace in detail the evolution of active galactic nuclei \citep{civano12, vito14}. 
Through the addition of data from the less deep \textit{XBootes, XMM-XXL} X-ray surveys (typical 0.5--2~keV fluxes $\sim10^{-14}$~erg/s/cm$^2$) the sky coverage area increases by a factor of $\sim 10$  \citep{ueda14,aird15,georgakakis15}. In their recent \cite{kalfonzou14} paper
constructed the X-ray luminosity function for quasars at $z>3$ in a field $\simeq 33$~sq.~deg. The survey of such a field was composed of the archival data from individual pointings of the
\textit{Chandra} satellite over the entire timeof its operation. \citep{kalfonzou14} managed to exclude some models of the luminosity function using the survey data, but the size of this sample turns out to be insufficient to investigate the properties of the population of bright (luminosities $>5\times 10^{44}$ erg/s) and distant ($z>3.5$) quasars. 

The data from the \XMM \ X-ray telescope
accumulated over 15 years represent a serendipitous X-ray sky survey \citep{watson09} with
a total area of $\sim$800~sq.~deg. and a sensitivity $\approx 5\times 10^{-15}$~erg\,s$^{-1}$\,см$^{-2}$ \citep[the \3XMMDR4 \ version \footnote{\url{http://heasarc.gsfc.nasa.gov/W3Browse/xmm-newton/xmmssc.html}},][]{watson09}. A sample of quasars at $z>3$ selected by their X-ray emission that exceeds the existing samples of \cite{kalfonzou14, georgakakis15} by several times can be obtained from the data of this survey.
 
Previously \citep{khorunzhev16}, we made an attempt to find new sources and to obtain a
more complete sample of X-ray quasars at $z>3$ in the fields of the serendipitous \XMM \ \3XMMDR4 \ survey at Galactic latitudes |b|>20$^\circ$
using photometric data from the Sloan Digital Sky Survey \citep[\SDSS,][]{alam15}, \textit{\2MASS} \citep{cutri03} and \textit{\WISE} \citep{wright10}. The area of the overlap between \3XMMDR4 \ and \SDSS is $\sim$300~sq.~deg. Based on the broadband photometry
of the above surveys, we obtained the photometric
redshift estimates ($\zphot$) using the
 \EAZY  software \citep{brammerdokkum08}. We compiled a catalog of
 903 candidates for distant quasars selected by the
photometric redshift (\K16). Both already known quasars
(with spectroscopic redshifts $\zspec>3$) and new unstudied objects
(with photometric redshift estimates $\zphot>2.75$) enter into the catalog.
\begin{figure*}[!ht]
\centering
\includegraphics[width=0.45\linewidth]{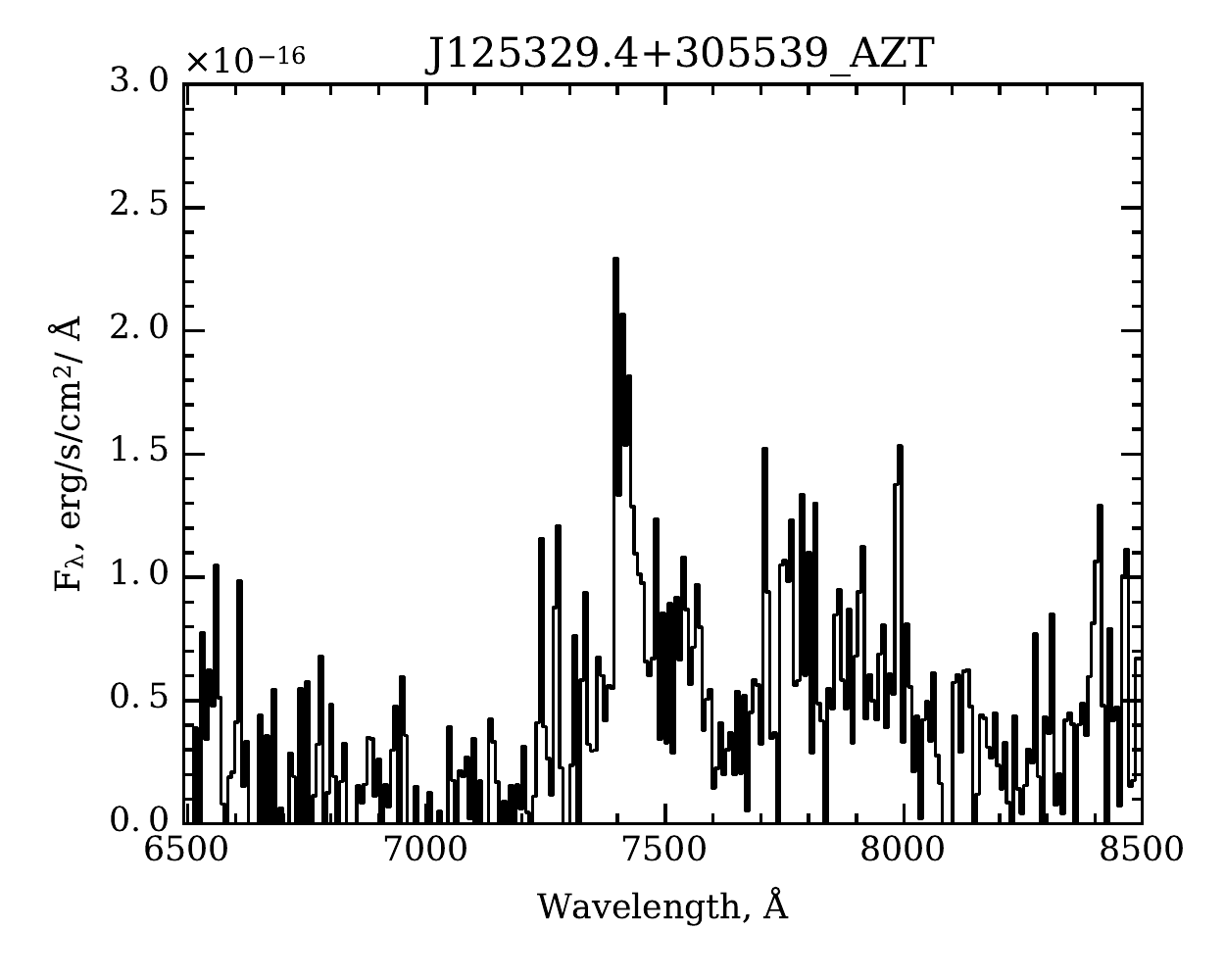}
\includegraphics[width=0.45\linewidth]{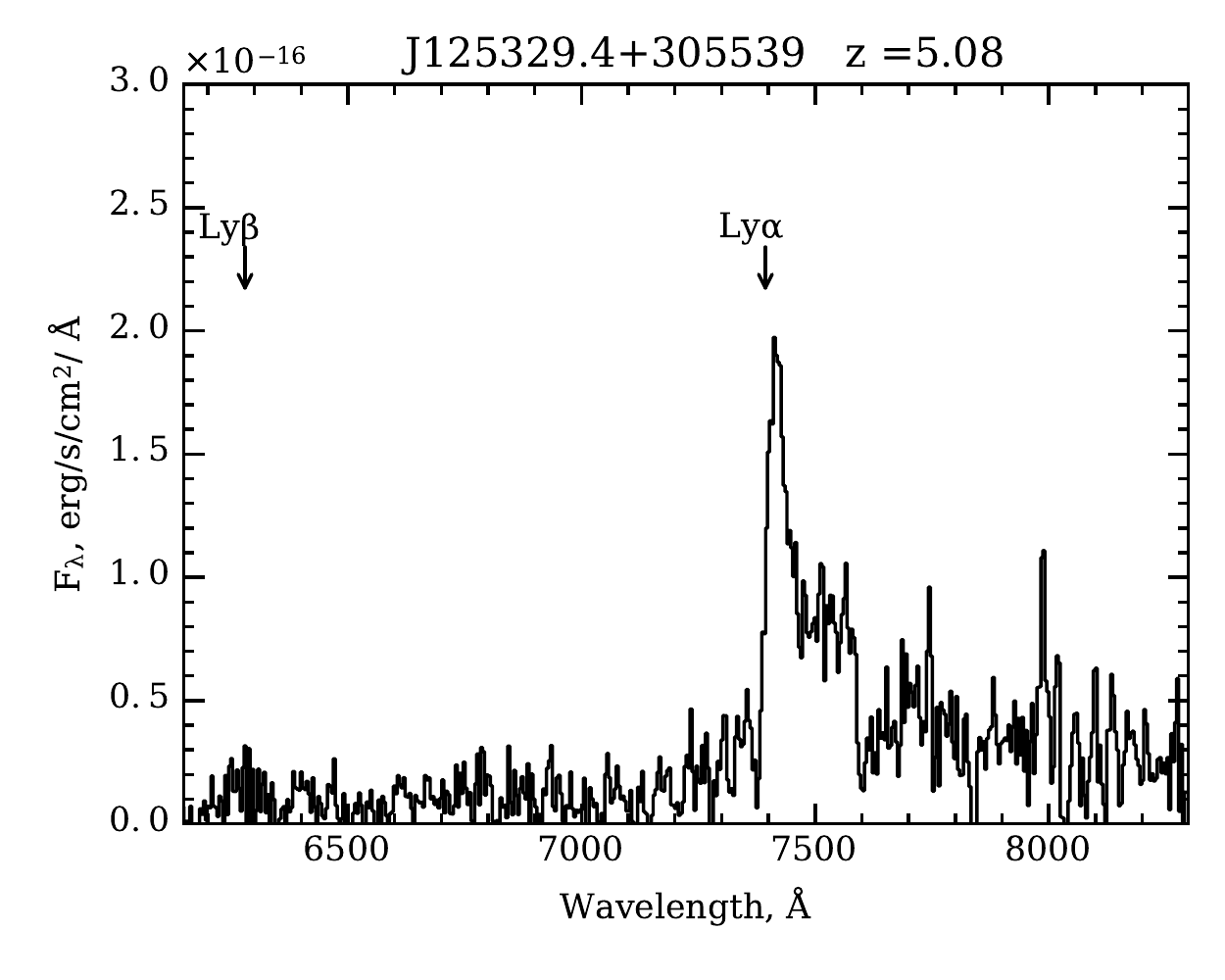}
\caption{Spectra of the distant quasar \textit{3XMM~J125329.4+305539} ($\zspec=5.08$) taken at the 1.6-m \Azt \ telescope
(left) and the 6-m BTA telescope (right). The neighboring spectral channels are binned by two along the wavelength axis.}
\label{fig:QSO5} 
\end{figure*}

The completeness of \K16 in the investigated fields
relative to the spectroscopic catalogs of quasars (\SDSS \citep{alam15} and \textit{The Half Million Quasars} \citep{flesch15}) with $\zspec > 3$  is about 80\%. The normalized median absolute deviation ($\Delta z = | \zspec - \zphot |$) is $\sigma_{\Delta z/(1+\zspec)} = 0.07$, while the outlier fraction is $\eta = \Voutl \%$ (when $\Delta z /(1+\zspec)>0.2$). 

In the \K16 catalog about 40\% of the candidates have no spectroscopic redshifts. These are new
quasar candidates. If most of them are actually quasars at $z > 3$, then the existing
 \3XMMDR4 \ sample of distant quasars can be expanded by a factor of $\sim 1.5$ \citep{khorunzhev16}. 
Spectroscopic verification is needed to understand the accuracy of the $\zphot$ estimate and to estimate the purity of the selection of quasar candidates. Obtaining the spectra of several
hundred objects “scattered” over the celestial sphere
is a laborious task. However, we can take the spectra
of a small “random” sample and draw preliminary
conclusions about the quality of the $\zphot$ estimates for
new sources based on them. We prepared a program
of observations of candidates for distant quasars at $3<z<5.5$  from the \K16\ catalog at the 1.6-m
 \Azt\ and 6-m BTA telescopes. By now the spectra of \Nobs \ objects have been taken as part of the program.
Below we present the results obtained from these data.

\section{2. OBSERVATIONS AT THE AZT-33IK TELESCOPE WITH THE LOW- AND MEDIUM-RESOLUTION ADAM SPECTROGRAPH}

A quasi-random spectroscopic survey of 18
quasar candidates from the \K16 \ catalog that previously had no spectroscopic redshifts has been
conducted at the \Azt \ telescope \citep{kamus02} equipped with the low- and medium-resolution
ADAM spectrograph \citep{afanasev16,burenin16}.

The \Azt \ telescope is located at the Sayan
Solar Observatory of the Institute of Solar–Terrestrial
Physics, the Siberian branch of the Russian Academy
of Sciences, and has a primary mirror diameter of
1.6 m. The ADAM spectrograph was produced at the
Special Astrophysical Observatory and was installed on \Azt \ in 2015. The main structural components of the spectrograph are: \textit{Andor Newton 920} array with an efficiency of $\sim$90\% in the range from
4000 to 8500\AA\ and a set of dispersive elements
(volume phase holographic gratings). The quantum
efficiency of the entire system (the telescope mirror,
the spectrograph, the grating, and the CCD array)
reaches 50\% \citep{burenin16}.

The target objects of the \K16 sample are quasars
with broad emission lines. The typical apparent magnitude of the objects is $i^\prime \sim 20.5$. 
The exposure time
was chosen to be sufficient for the bright emission
lines from which the quasar redshift could be determined 
to manifest themselves. This allows the
spectra for quite a few sources to be taken with a small
telescope. Spectra with a higher signal-to-noise ratio
in continuum are required to determine the redshift
and type of sources without bright lines. Repeated
observations with a longer exposure time or at larger
telescopes are needed for this purpose.

The properties of the \K16 \ sample change significantly 
with increasing redshift: the number of objects drops exponentially; 
the X-ray and optical fluxes
become fainter. On average, the \K16 objects are at
 $\zspec\sim3$ and have magnitudes
 $i^\prime \sim 20.5$. There are
only a few dozen candidates at $\zphot > 4$ with magnitudes $i^\prime >20.5$ in the
 \K16 \ catalog. We selected the
sources for our observations almost randomly within
two ranges:
$2.75<\zphot<4$ and $\zphot\geqslant4$. However, in the range $2.75<\zphot<4$ we preferentially
observed brigh candidates with  $i^\prime < 20$. These peculiarities of 
the selection of objects for our spectroscopic program 
should be taken into account when
formulating the conclusions about the purity of the
entire \K16 catalog of quasars.

The quasar candidates were observed in dark time
(the lunar phase is less than 0.3) and at a mean
seeing better than 2-arcsec. Under such conditions,
a one-hour exposure time is sufficient for the detection 
of emission lines in the spectra of quasars
with magnitudes $i^\prime = 20.5$.
For our observations we
used a 2-arcsec-wide slit. The objects at
 $\zphot <3.5$, $3.5<\zphot<4.5$, $4.5<\zphot$  were observed with the
 VPHG600G (the range is 3700--7340\AA, the resolution is 8.8\AA), VPHG300 (the range is 3900--10500\AA, the resolution is 13.8\AA), and VPHG600R (the range is  6520--10100\AA, the resolution is 7.3\AA) gratings, respectively.
The above resolutions were
achieved for the 2-arcsec-wide slit. We chose such
a grating that the presumed position of the
 Ly$\alpha$ line
was near the peak of its diffraction efficiency. The data
were reduced with the standard IRAF\footnote{http://iraf.noao.edu} software.

\section{3. Results}
The list of objects is given in Table 1 (end of this paper, their
spectra are shown below in Fig. 3). The shape of
the spectra was corrected using the observations
of spectrophotometric standards from the list by
Massey et al. (1988).
Fourteen of the \Nobs \ are
quasars; 10 of them are quasars at $\zspec >3$. Their
redshifts were determined from the positions of the
peaks of broad lines in the spectrum. The types of the
remaining objects are difficult to determine, because
there are no bright emission lines in their spectra. 

The accuracy of the redshift for distant objects
depends on the spectrograph resolution as $(1+z)\times \times\frac{\Delta \lambda}{\lambda}$ and is approximately 0.01
for low-resolution spectra.
Therefore, the spectroscopic redshifts for the objects
are given to the second decimal place. The shape
and positions of broad lines are known to be closely
related to the processes occurring near a black hole.
The redshift determined from broad lines can slightly
differ from $\zspec$ of the host galaxy. The redshifts in
the spectra where only the $Ly \alpha$ line is seen should
be treated with caution. Its shape can be severely
distorted by absorption and, consequently, the position
of its peak can be determined incorrectly. Such
objects are marked by the quality flag (QF) = 1 in
Table 1.

The spectroscopic sample of \Nobs \ “randomly” selected objects has a median 0.5-2 keV X-ray flux
$\simeq 5 \times 10^{-15}$ erg/s/cm$^2$. This value coincides with
the median X-ray flux of the \K16 \ sources. Among
the selected objects there are no bright quasars with
strong emission lines at $\zspec>3$ with 0.5-2 keV fluxes $> 10^{-14}$ erg/s/cm$^2$. The median apparent
magnitude is $i^\prime = 19.9$, which is brighter than the
mean value for the \K16 \ by 0.5 magnitude. Thus, the sample of \Nobs \ sources may be deemed representative 
in X-ray flux for the \K16, but not in optical flux.

\subsection{3.1. The Quasar 3XMM~J125329.4+305539 at $z=5.08$ }

The distant X-ray quasar \textit{3XMM~J125329.4+305539} at $\zspec=5.08$ and with 
an apparent magnitude $i^\prime=21$ was discovered and confirmed at the \Azt \ and BTA telescopes.
The \Azt \ and BTA spectra are shown in Fig.~\ref{fig:QSO5}.

The first spectrum of this object was taken at the \Azt \ telescope with the ADAM spectrograph
with an exposure time of 1.5~h. From the spectrum
we managed to determine that the source is a distant
quasar and to measure its redshift $\zspec~=~5.1$. 
Telescopes with a larger diameter are usually required to
take the spectra of such sources. However, as can
be seen from our results (see Fig.~\ref{fig:QSO5}), using the new
spectrograph with a high quantum efficiency in the
near infrared at the 1.6-m \Azt \ telescope, we can take the spectra of faint objects (down to $i^\prime \simeq$21), and determine their types and redshifts.

To improve the spectroscopic redshift, we took a
spectrum with a higher signal-to-noise 
ratio and a resolution of 18\AA \ at the 6-m BTA telescope using
the SCORPIO spectrograph \citep{afanasev05} with an exposure time of 0.5 h. 
The redshift $\zspec = 5.08\pm0.01$ was determined by fitting the
spectrum by the template of a type~1 quasar
 \citep{vandenberk01} into which the interstellar absorption
by neutral hydrogen was introduced
 \citep{madau95}. The error of the template position is underestimated,
because the template is an averaged model of the
spectrum that disregards the individual deviations in
the spectrum of the separate source. Therefore, the
error was estimated via the spectral resolution of the
instrument.

\textit{3XMM J125329.4+305539} was first reported
as a probable quasar at $\zphot=4.64$ in the \K16 catalog, and there was no information about this
source in other photometric catalogs of quasar candidates\footnote{\url{http://vizier.u-strasbg.fr}}. 
It was not an XMM-Newton target (i.e., it fell within its field of view by chance). 
Its SDSS apparent magnitude is $i^\prime$ $\simeq 21.0$. The 0.5--2 keV X-ray flux is $1.5\times \times10^{-15}$~erg/s/cm$^2$, the 0.5--2~keV luminosity is $4 \times10^{44}$~erg/s in the observer's frame.

Apart from this source, only three optically bright
(there is reliable SDSS photometry) X-ray quasars
at $\zspec>5.0$ that were not target objects for pointing \XMM
are known in the \3XMMDR4 \ catalog: \textit{3XMM J221643.9+001346} ($\zspec=5.01$, $i^\prime \simeq 20.3$) \cite{anderson01,gavignaud06}, \textit{3XMM J011544.8+001513} ($\zspec=5.10$, $i^\prime \simeq 21.4$) \cite{mcgreer13},  \textit{3XMM J022112.5-034251} ($\zspec=5.01$, $i^\prime \simeq 19.3$) \cite{paris16}. 
The first two sources were found in the \SDSS\ \textit{Stripe~82},
where the \SDSS\ sensitivity is much better than that averaged over the sky and, 
consequently, the selection completeness is higher.
 
Note that there is one more source in the \K16 \ catalog,
\textit{3XMM J004054.6-091527}, with the published
spectroscopic redshift $\zspec$=5.002 from \SDSS \ data release 12 (DR12). However, fresher results of spectroscopy
($\zspec$=4.980$\pm$0.010) for this source obtained at a telescope with a larger diameter are presented in
\citep{worseck14}. Therefore, \textit{3XMM J004054.6-091527} is no longer considered as a quasar at
$\zspec>5.0$.

Thus, the object \textit{3XMM J125329.4+305539} investigated by us is one of the brightest and most
distant X-ray quasars at $\zspec>5.0$ suitable for 
constructing the X-ray luminosity function at such redshifts. 
The redshifts, magnitudes, and X-ray fluxes
of \textit{3XMM J125329.4+305539} and the other three
distant quasars listed above are given in Table 2.
Note that there are also even more distant quasars in the
 \3XMMDR4 \ catalog, but they were target sources
for pointing the X-ray telescope (after their discovery
in the optical band) and, therefore, cannot be used in
constructing the X-ray luminosity function.

\begin{table}
\small
\caption*{\textbf{Table 2.} Properties of the selected X-ray quasars at $\zspec >5$.}
\footnotesize
\begin{tabular}{lccccc}
\hline
\hline
Name (XMM) & $\zphot$ & $\zspec$ & $i^\prime$ & $F^{-14}$ & L$_{0.5-2}$ \\
\hline
J011544.8+001513 & 0.61 & 5.10 & 21.4 & 0.19 & 44.7 \\
J022112.5$-$034251 & 4.74 & 5.01 & 19.3 & 0.61 & 45.2 \\
\textbf{J125329.4+305539} & 4.64 & 5.08 & 20.9 & 0.15 & 44.6 \\
J221643.9+001346 & 4.91 & 5.01 & 20.3 & 0.22 & 44.8 \\
\hline
\hline
\end{tabular}
\small
\textbf{Note.} $\zphot$ is the photometric redshift of the object in the K16 catalog; $\zspec$ is the spectroscopic
redshift; $i^\prime$ is the apparent magnitude in the  SDSS $i^\prime$ band; $F^{-14}$ is the 0.5--2~keV X-ray flux (erg/s/cm$^2$) normalized to  $10^{-14}$; $L_{0.5-2}$ is the common logarithm of the X-ray luminosity (erg/s in 0.5--2~keV in the observer's frame).
\end{table}

\subsection{3.2. Remarks on Infividual Objects}
\par{3XMM J025459.8+192343}. This source ($\zspec=2.81$) does not enter into the published 
 \K16 \ catalog, but, nevertheless, it was included in Table~1. Its spectrum was taken on the first nights of ADAM 
  operation, when it was a candidate in the intermediate version of \K16. 
We thought that the photometric redshift of the source was  $\zphot=2.6$, and it 
could theoretically be at  ${\zspec>3}$. This source enters 
into the catalog of quasar candidates by \citep{richards15}, 
where its best photometric redshift estimate is 3.295.

\par{3XMM J062923.4+634935}. There is a set of 
narrow absorption lines in the spectrum of this quasar 
 ($\zspec=2.88$). We assume that the most distinct lines are
$\lambda_{H\beta 4861} = 5696$\AA, $\lambda_{MgI 5175}=6077$\AA, and $\lambda_{NaI 5891}=6895$\AA. 
A cloud of intergalactic gas that
gives such a line structure probably lies on the line of
sight between us and the object (at $\zspec \simeq 0.17$).

\par{3XMM J103901.4+643335}. This is a distant
quasar at $\zspec = 4.08$. It is located in the region of
the overlap between the \XMM \ and \textit{Chandra} surveys. Both telescope detected an X-ray flux from
this source. The source was first reported as a
quasar candidate in the \K16 \ with ${\zphot=4.01}$. There are only a few dozen known X-ray
quasars at $\zspec \sim 4$. Therefore, the confirmation of
this source has a high significance for studying the
the population of quasars at such redshifts.
    
\par{3XMM J131213.0+352347}. This source has
 $\zphot=4.92$. Feature characteristic of M-type stars
are seen against the background of big noise in the
spectrum. 

\subsection{Additional Remarks to the K16 Catalog}
After the publication of the K16 catalog, having 
additionally browsed the literature, we found
that two X-ray sources, \textit{3XMM J122004.8+291304} and \textit{3XMM J172014.1+264712}, 
were erroneously
included in the K16 catalog as quasar candidates.
These X-ray sources turned out to be objects of the
nearby Universe.

\par{3XMM J122004.8+291304}. This is a globular
cluster in the halo of the nearby galaxy NGC 4278 (\textit{NGC 4278-X30} or \textit{CXO J122005.011+291304.73} \citep{liu2011,usher12}). 
\par{3XMM J172014.1+264712}. The source is the
cluster galaxy \textit{RX J1720.1+2638} at $z=0.338$ \citep{owers11}.

This information is included in Table 1.
 
\subsection{3.4. Purity of the Quasi-Random Spectroscopic Sample and the K16 Catalog}
\begin{figure}
\includegraphics[width=0.95\linewidth]{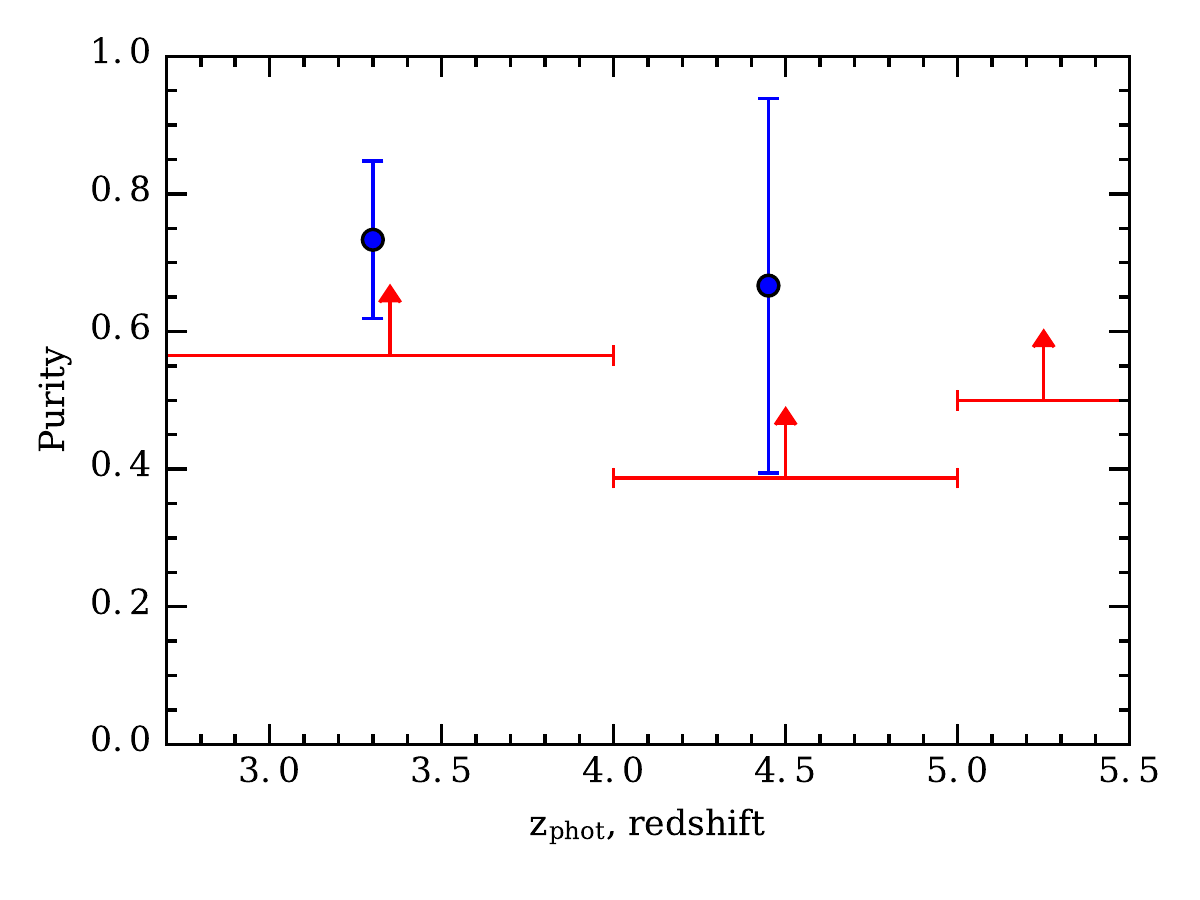}
\caption{The circles with Poissonian errors indicate the purity of the 
\Nobs \ quasar candidates whose spectra were taken at \Azt. The arrows indicate the 
extimated lower limit for the purity of the \K16 catalog relative to the objects 
with known (from literature or SDSS) spectroscopic redshifts.
}
\label{fig:Purity} 
\end{figure}

We determined the spectroscopic redshifts of the \Nobs \ quasar candidates selected quasi-randomly from
the \K16 catalog.
Let us estimate the purity of
this sample in wide ranges of photometric redshifts:
 $2.75\leq \zphot <4$, $4\leq \zphot <5$, $5\leq \zphot <5.5$. By the purity we mean the ratio of the number of
true quasars (\Zphzsp$<0.2$) to the
number of all objects with the available spectra. The
condition \Zphzsp$<0.2$ is introduced to take into account the scatter of
 $\zphot$ relative to $\zspec$. A value of 0.2 roughly corresponds to three
standard deviations of
$\zphot$ relative to $\zspec$ for all of
the known and spectroscopically confirmed quasars
 from the complete \K16 \ catalog (see~\cite{khorunzhev16}). The purity of the spectroscopic sample
of \Nobs \ objects calculated in this way is shown in Fig.~\ref{fig:Purity} (circles). 

For comparison, the arrows in Fig.~\ref{fig:Purity} indicate the
lower limit for the purity of the entire \K16 \ catalog (without the \Azt observations). 
This limit was deduced as the ratio of the number of true
quasars with known spectroscopic redshifts and \Zphzsp$<0.2$ to the total number
of objects in the catalog. Recall that the candidate
sources without spectroscopic redshifts accounted for
about 40\% of the \K16 objects. 

Since the purity of the \K16 catalog was initially
higher than 50\% and since more than half of the \Nobs \ objects with optical \Azt \ spectroscopy turned out
to be quasars at $\zspec >3$, we can draw the preliminary conclusion 
that the true purity of the K16 catalog
of candidates for distant quasars is
 70--80\%. This conclusion is yet to be refined, because the sample for
our observations at the \Azt \ elescope was not
absolutely random and consisted of relatively bright
(for the \K16) objects.

\section{4. Conclusions}

In this paper we showed that the purity of the
quasi-random sample of 18 quasar candidates from
the \K16 catalog exceeds 50\%. Strictly speaking, this
is true only for optically bright sources, $i^\prime \lesssim 20$. We
are going to continue the spectroscopy of candidates
from the K16 sample with the ADAM spectrograph
at the \Azt \ telescope. However, a telescope
with a larger diameter will be required to check the
bulk of the \K16 objects
($i^\prime \sim 20.5$, $\zphot \sim 3$). It is
expected that faint and distant ($\zphot>4$) objects will
be observed at the 6-m BTA telescope.

Our spectroscopy of the selected X-ray candidates 
for distant quasars detected by an improved
selection method based on publicly accessible SDSS
and WISE photometric data confirmed that more
quasars than in the available catalogs could be found
 \citep{richards15,dipompeo15}. 
The discovery of one of the most distant selected X-ray
quasars (3XMM~J125329.4+305539 at $\zspec=5.08$) proves this convincingly.
 
Apart from the new quasar 3XMM~J125329.4+305539 the K16 catalog contains three more 
optically bright ($i^\prime<21$) quasars at $\zspec >5$. 
The X-ray fluxes and luminosities of these four objects in the
0.5--2~keV energy band exceed $1.5 \times 10^{-15}$~erg/s/cm$^2$
and $4\times 10^{44}$~erg/s, respectively. About 50~sq.~deg. is covered with this or
better sensitivity in the regions of the overlap between \3XMMDR4\ and \SDSS (see Fig.~9 in \citealt{khorunzhev16}). Therefore, it is interesting to note that
approximately the same ($\sim3\times 10^{-15}$~erg/s/cm$^2$) limiting sensitivity 
must be achieved in the planned
four-year sky survey by the eROSITA telescope
onboard the SRG observatory near the ecliptic poles
in a field of about 150~sq.~deg. \citep{merloni14}. 
Consequently, it will be possible to detect several
new optically bright quasars at $z >5$ with an X-ray
luminosity higher than
 $\sim 10^{45}$~erg/s in these fields. 
At the same time, the total number of quasars at
 $z >5$ discovered by the eROSITA telescope near the
ecliptic poles may turn out to be much greater (tens
or even hundreds; \citealt{kolodzig13}), but most
of them will most likely be fainter than the SDSS
sensitivity threshold. Deeper optical surveys, for
example, \textit{PanSTARRS} \cite{hodapp04} and the
sky surveys conducted by the
 \textit{Hyper Suprime-Cam} \citep{miyazaki12} at the 8-m Subaru
telescope, will be required for their identification.
Note also that the \K16 \ catalog has no X-ray quasars
or quasar candidates at $z>5$ with fluxes higher than
 $10^{-14}$~erg/s/cm$^2$ (corresponds to the mean all-sky
sensitivity of the four-year eROSITA survey), 
i.e., a 0.5--2~keV luminosity higher than $\sim 3\times 10^{45}$~erg/s in a field of $\simeq 250$~кв.~град. This is consistent with
the predictions made by \cite{georgakakis15} based on their model of the X-ray luminosity function 
for quasars at
at $3<z<5$. Consequently, one might
expect no more than $\sim 500$ optically bright quasars at
 $z>5$ with a luminosity higher than
 $3\times 10^{45}$~erg/s (0.5--2~keV) to be found in the eROSITA survey.

Our quasar spectra demonstrate the unique 
capabilities of the new ADAM spectrograph installed on
the  \Azt \ telescope at the Sayan Solar Observatory.
The ADAM spectrograph allows the spectra
of objects with an apparent magnitude R$\sim$19.5 
to be taken with an exposure time of 30 min.
If necessary and under good weather conditions, 
a magnitude I$\sim$21 can be reached with an exposure time 
of two hours.
The ADAM spectrograph is planned to be one
of the instruments for the optical support of the SRG
project \citep{merloni14,pavlinsky11}.
   
\paragraph{ACKNOWLEDGMENTS}  
This study was supported by the Russian Science
Foundation (project no. 14-22-00271 / RSF 14-22-00271). 
The observations at the 6-m BTA telescope were 
financially supported by the Ministry of 
Education and Science of
the Russian Federation
(contract no. №14.619.21.0004, project identifier
 RFMEFI61914X0004). 
We are
grateful to the referees for a number of valuable
remarks. 
We would like to thank V.~Astakhov for translation of 
the paper in English.
It have been used the VizieR catalogue access tool, CDS,
\citep{ochsenbein00}.

\vfill
\eject

\onecolumn

\begin{landscape}
\begin{center}
\textbf{Table 1.} Redshifts for the quasi-random spectroscopic samle
\end{center}

\begin{longtable}{ccD{.}{.}{3}D{.}{.}{3}cccD{.}{.}{2}D{.}{.}{3}cccc}

\hline
\hline
\multicolumn{1}{c}{Name 3XMM} &
\multicolumn{1}{c}{Date} &
\multicolumn{1}{c}{RA} &
\multicolumn{1}{c}{DEC} &
\multicolumn{1}{c}{OBJID SDSS} &
\multicolumn{1}{c}{$F^{-14}_{0.5-2}$} &
\multicolumn{1}{c}{$i^\prime_{PSF}$} &
\multicolumn{1}{c}{$\zphot$} &
\multicolumn{1}{c}{$\zspec$} &
\multicolumn{1}{c}{QF} &
\multicolumn{1}{c}{$\zphot$$_{D15}$} &
\multicolumn{1}{c}{$\zphot$$_{R15}$} &
\multicolumn{1}{c}{L$_{0.5-2}$}
\\
\hline
\multicolumn{1}{c}{(1)} &
\multicolumn{1}{c}{(2)} &
\multicolumn{1}{c}{(3)} &
\multicolumn{1}{c}{(4)} &
\multicolumn{1}{c}{(5)} &
\multicolumn{1}{c}{(6)} &
\multicolumn{1}{c}{(7)} &
\multicolumn{1}{c}{(8)} &
\multicolumn{1}{c}{(9)} &
\multicolumn{1}{c}{(10)} &
\multicolumn{1}{c}{(11)} &
\multicolumn{1}{c}{(12)} &
\multicolumn{1}{c}{(13)}
\\

\hline
\endhead
    
\hline 
\endfoot
J025459.8+192343&2015/10/13&43.7490&19.3957&1237673283585769514&1.448&19.17&2.60&2.81&*$^1$&&&44.98\\
J062923.4+634935&2015/11/16&97.3468&63.8263&1237666462651646831&0.724&19.37&3.16&2.88&0&2.89&3.30&44.71\\
J074047.4+310856&2016/03/04&115.1979&31.1490&1237654627323216286&0.687&19.36&2.88&3.04&1&2.98&2.92&44.74\\
J074405.8+284354&2016/03/07&116.0246&28.7321&1237657119477924495&2.213&20.29&3.46&&1&&&\\
J091740.4+161412&2016/03/09&139.4182&16.2366&1237667782815777585&0.412&20.27&3.42&3.51&1&3.46&3.45&44.67\\
J103148.9+584418&2015/11/16&157.9533&58.7395&1237655368745222588&0.492&19.60&3.33&3.62&0&3.60&&44.78\\
J103901.4+643335&2016/03/05&159.7552&64.5593&1237651271895941447&0.332&20.55&4.01&4.08&1&&&44.73\\
J110518.4+250027&2016/02/04&166.3267&25.0085&1237667551956435181&0.377&19.95&3.21&3.56&1&3.51&&44.64\\
J114529.7+024647&2016/03/04&176.3740&2.7799&1237654030330691765&0.569&19.01&2.85&2.68&0&2.72&&44.52\\
J120641.1+651138&2016/02/05&181.6706&65.1941&1237651066815709744&0.178&19.53&3.34&3.47&0&&3.45&44.29\\
J124232.3+141729&2016/03/10&190.6350&14.2914&1237662524694528017&0.517&18.39&2.82&&1&1.91&&\\
J125329.4+305539&2016/03/09&193.3721&30.9277&1237667255629579190&0.155&20.99&4.64&5.08&0&&&44.62\\
J131213.0+352347&2016/03/07&198.0539&35.3966&1237665026520318032&0.029&20.52&4.92&&1&5.15&&\\
J133200.0+503613&2016/02/05&202.9998&50.6037&1237662301357736036&0.776&19.59&3.78&3.83&1&3.93&&45.03\\
J135538.5+383210&2016/03/08&208.9105&38.5361&1237662226223071466&0.395&19.23&2.90&2.86&1&&2.97&44.43\\
J141625.4+361901&2016/04/07&214.1057&36.3162&1237662225151361223&1.573&19.86&3.33&&1&&&\\
J151633.3+071039&2016/03/05&229.1385&7.1777&1237662237485564867&0.375&20.00&3.75&3.81&1&3.90&&44.71\\
J215240.0+140206&2015/10/13&328.1669&14.0351&1237678601291760405&2.155&19.45&3.15&2.17&0&2.38&&44.88\\
J234214.1+303606&2015/09/18&355.5590&30.6017&1237666183498039666&0.799&19.92&3.20&3.37&0&3.42&&44.91\\
\hline    
J122004.8+291304& &185.0209&29.2179&1237665440975159694&0.279&20.73&3.38&0.002&*$^2$&&&\\
J172014.1+264712& &260.0592&26.7864&1237655501891110649&0.436&20.96&4.21&0.338&*$^3$&&&\\

\hline
\hline

\end{longtable}
\blfootnote{
\textbf{Table 1.} Name --- is the name in 3XMM-DR4 (3XMMJ...), Date --- year/month/day when the first spectrum of the
object was taken, RA is the right ascension, DEC is the declination, OBJID SDSS is the unique number 
in the photometric SDSS catalog, $F^{-14}_{0.5-2}$ is the 0.5--2~keV X-ray flux (erg/s/cm$^2$) normalized to 10$^{-14}$, \\
{$i^\prime$ is the apparent} magitude in the SDSS  $i^\prime$ band (AB, PSF), 
$\zphot$ is the photometric redshift in the \K16 catalog, $\zspec$ is the spectroscopic redshift, 
QF is the quality flag for $\zspec$ (0 marks the redshift measured from several lines, 1 marks the redshift determined from only one $Ly\alpha$ line, *$^1$ means that the object does not enter into th \K16 catalog, 
*$^2$ means that the object is a globular cluster in a nearby galaxy \cite{liu2011,usher12}, *$^3$ means that the object is a galaxy \citep{owers11}),
$\zphot$$_{D15}$ is the photometric redshift (PEAKZ) in the catalog by \citep{dipompeo15},
$\zphot$$_{R15}$ is the photometric redshift (ZPHOTBEST) in the catalog by \citep{richards15}. 
L$_{0.5-2}$ --- is the common logarithm of the 0.5--2~keV X-ray luminosity}
\end{landscape}

\onecolumn
{\footnotesize
\textbf{Fig. 3.} Spectra of 19 quasar candidates (including 3XMM J025459.8+192343that does not enter into the final version of the \K16 catalog) taken with the ADAM spectrograph at the \Azt\ telescope or (only \textit{3XMM J125329.4+305539}) with the SCORPIO spectrograph at the BTA telescope.
For \textit{3XMM J062923.4+634935} the horizontal line with marks indicates
the set of absorption lines in the cloud of intergalactic gas at
$\zspec \simeq 0.17$: $\lambda_{H\beta 4861} = 5696$\AA, $\lambda_{MgI 5175}=6077$\AA, $\lambda_{NaI 5891}=6895$\AA. 
The neighboring spectral channels were binned by two along the wavelength axis.

\label{fig:spimage} 
}
\begin{center}

\includegraphics[width=0.45\linewidth]{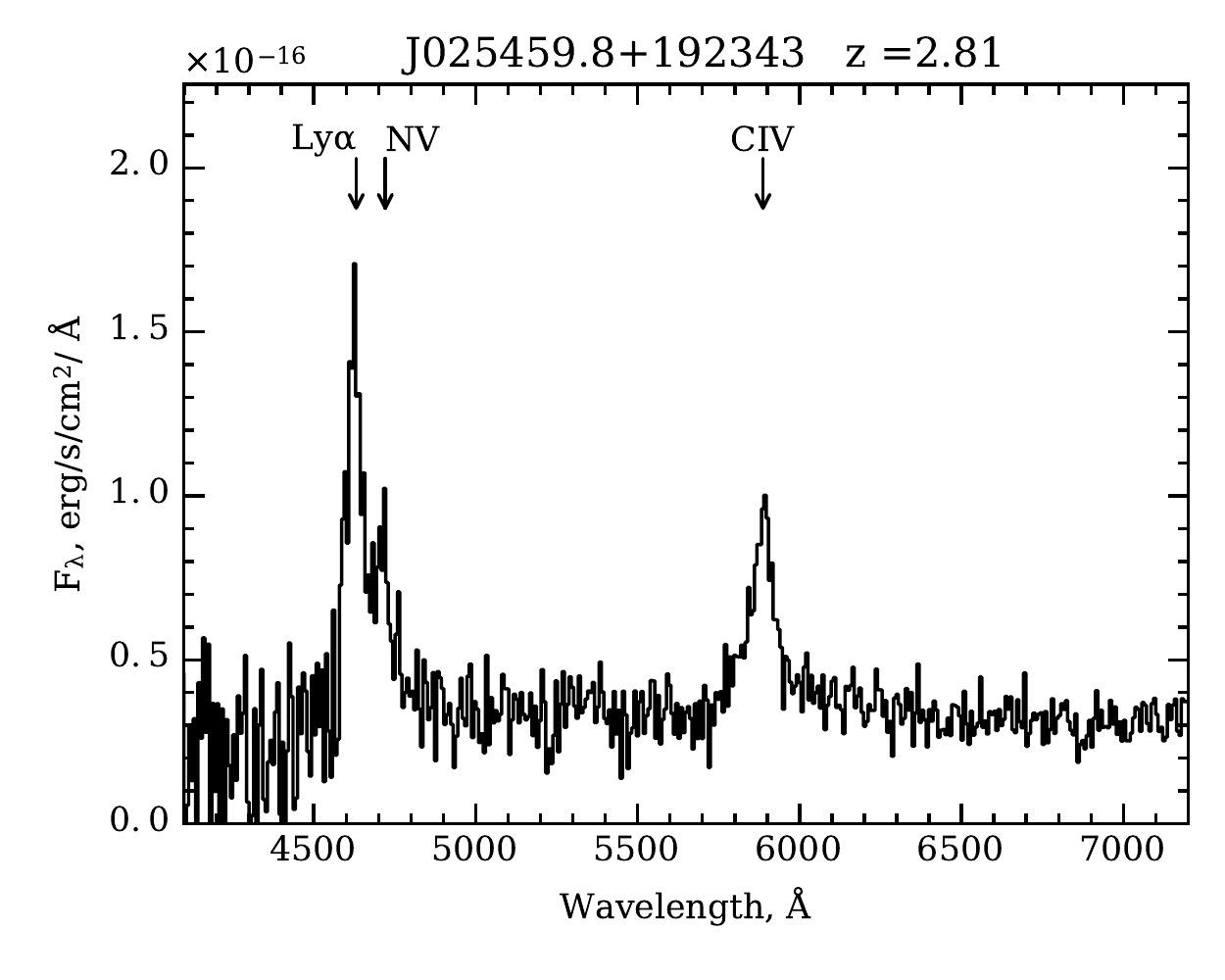}
\includegraphics[width=0.45\linewidth]{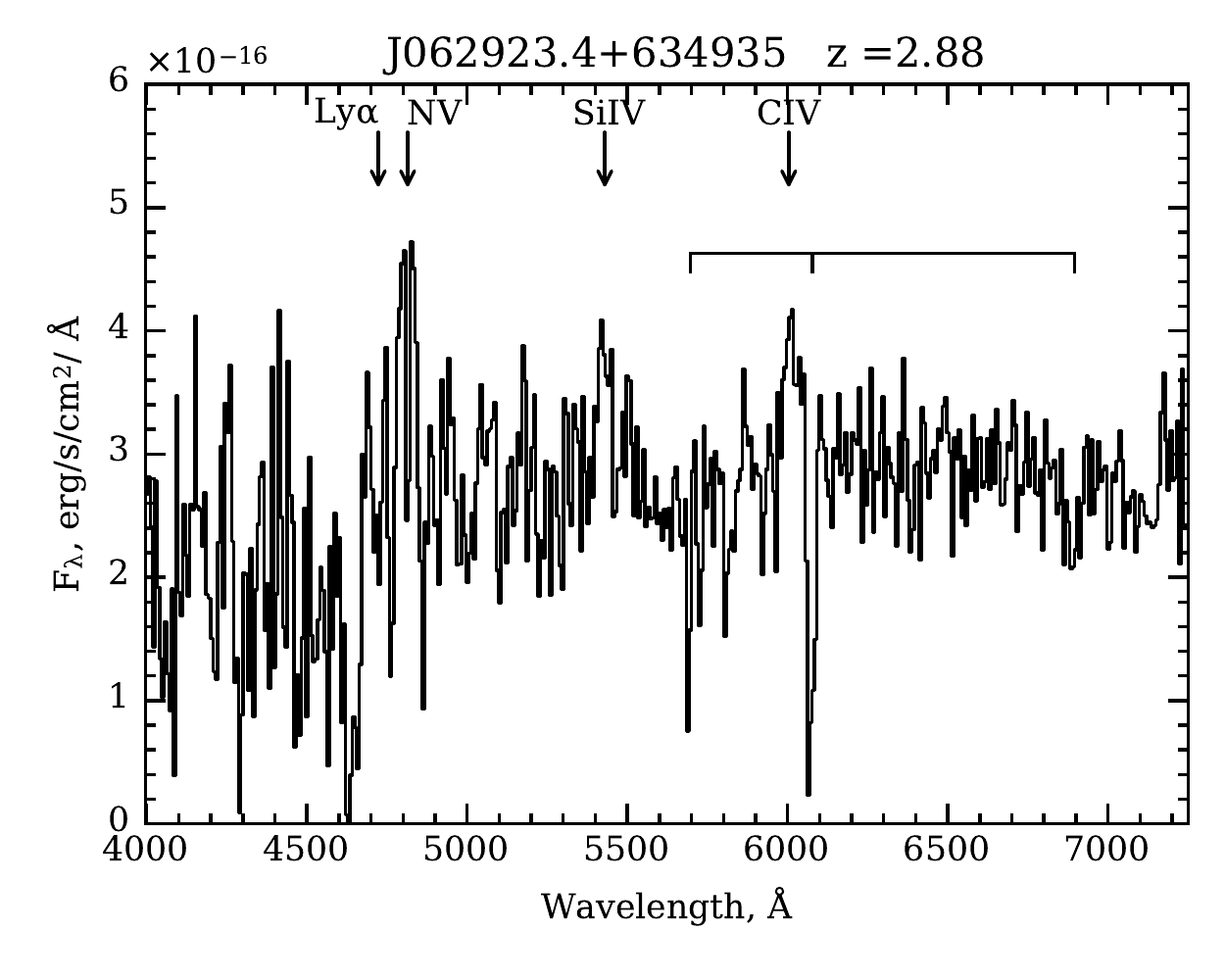}
\includegraphics[width=0.45\linewidth]{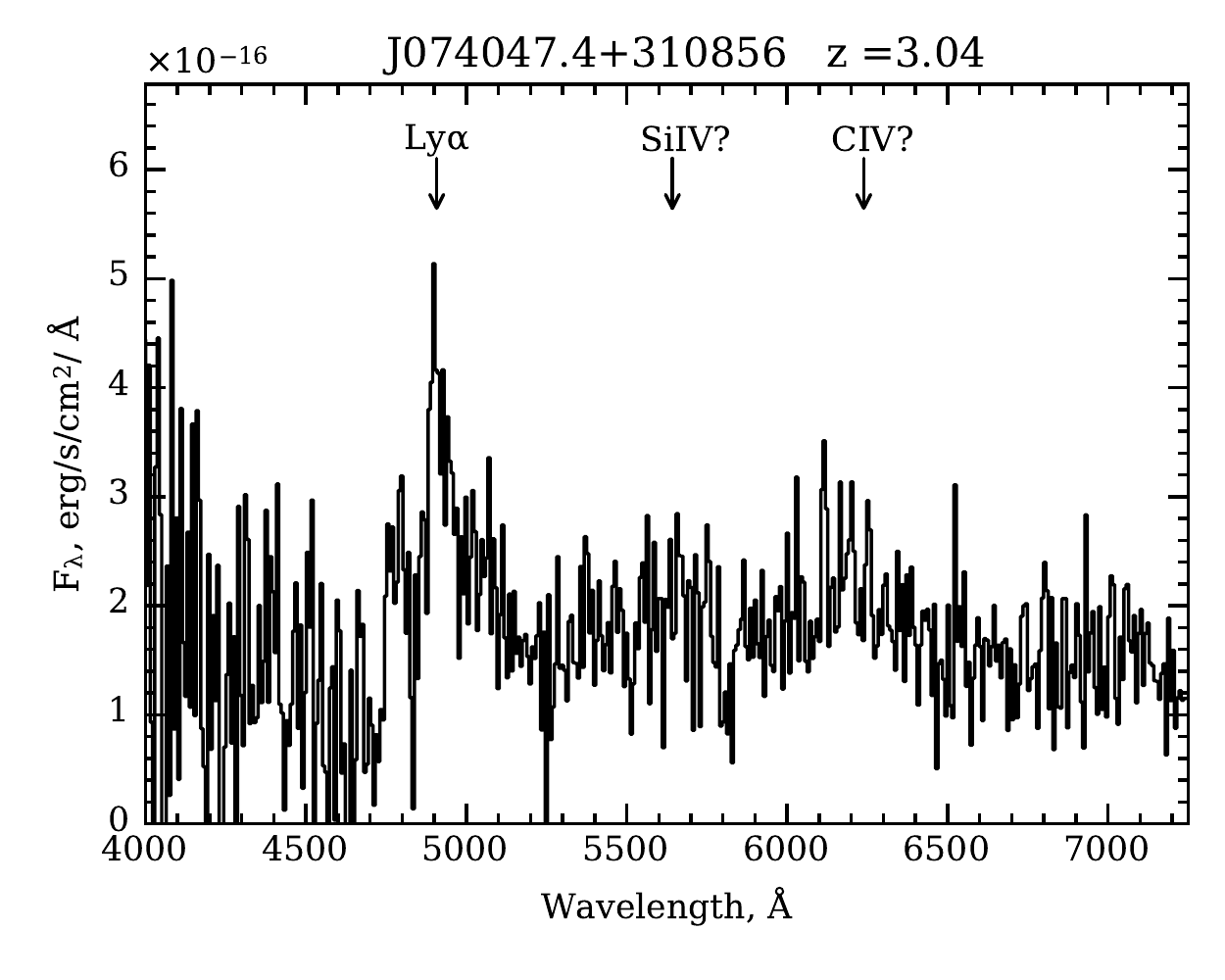}
\includegraphics[width=0.45\linewidth]{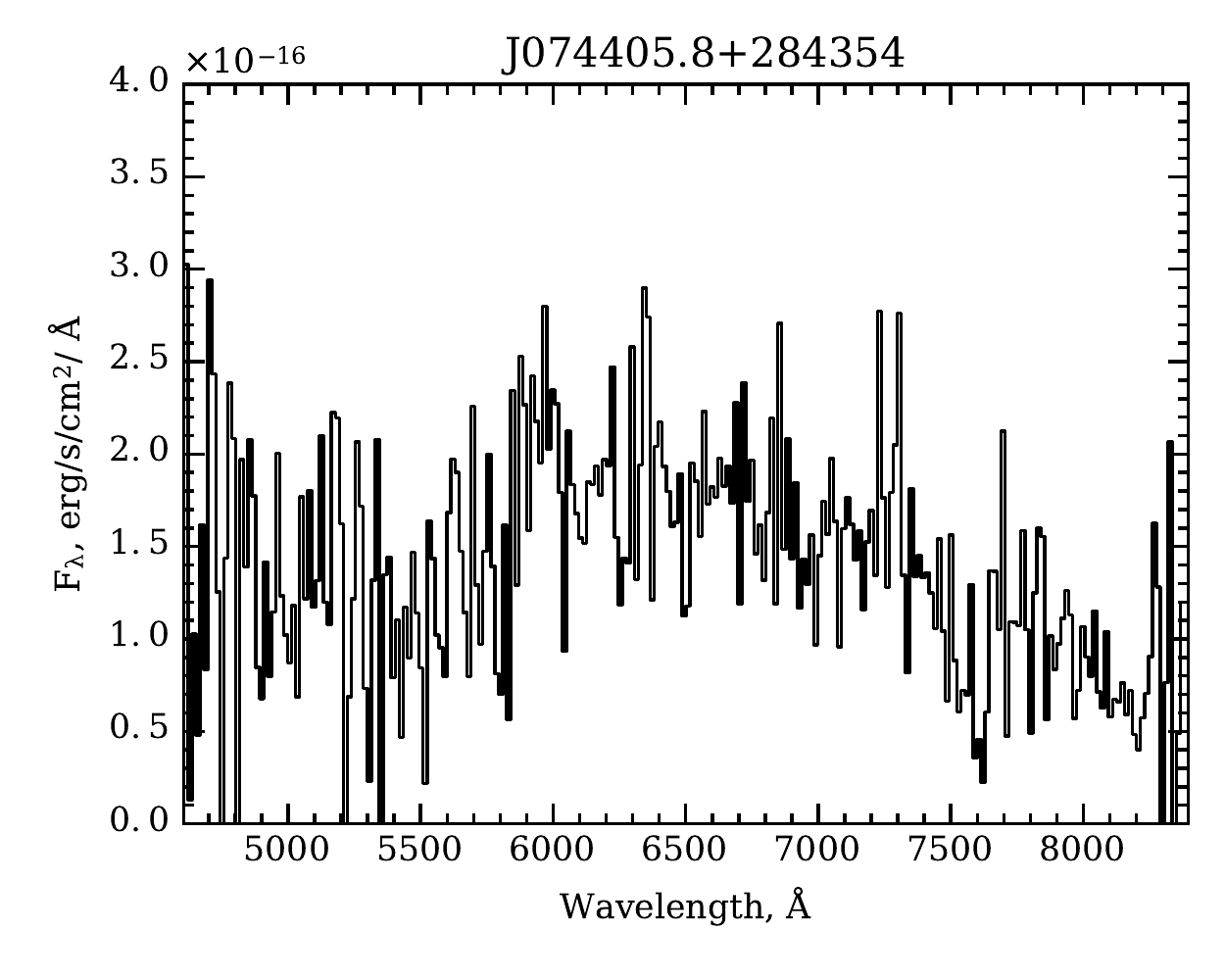}
\includegraphics[width=0.45\linewidth]{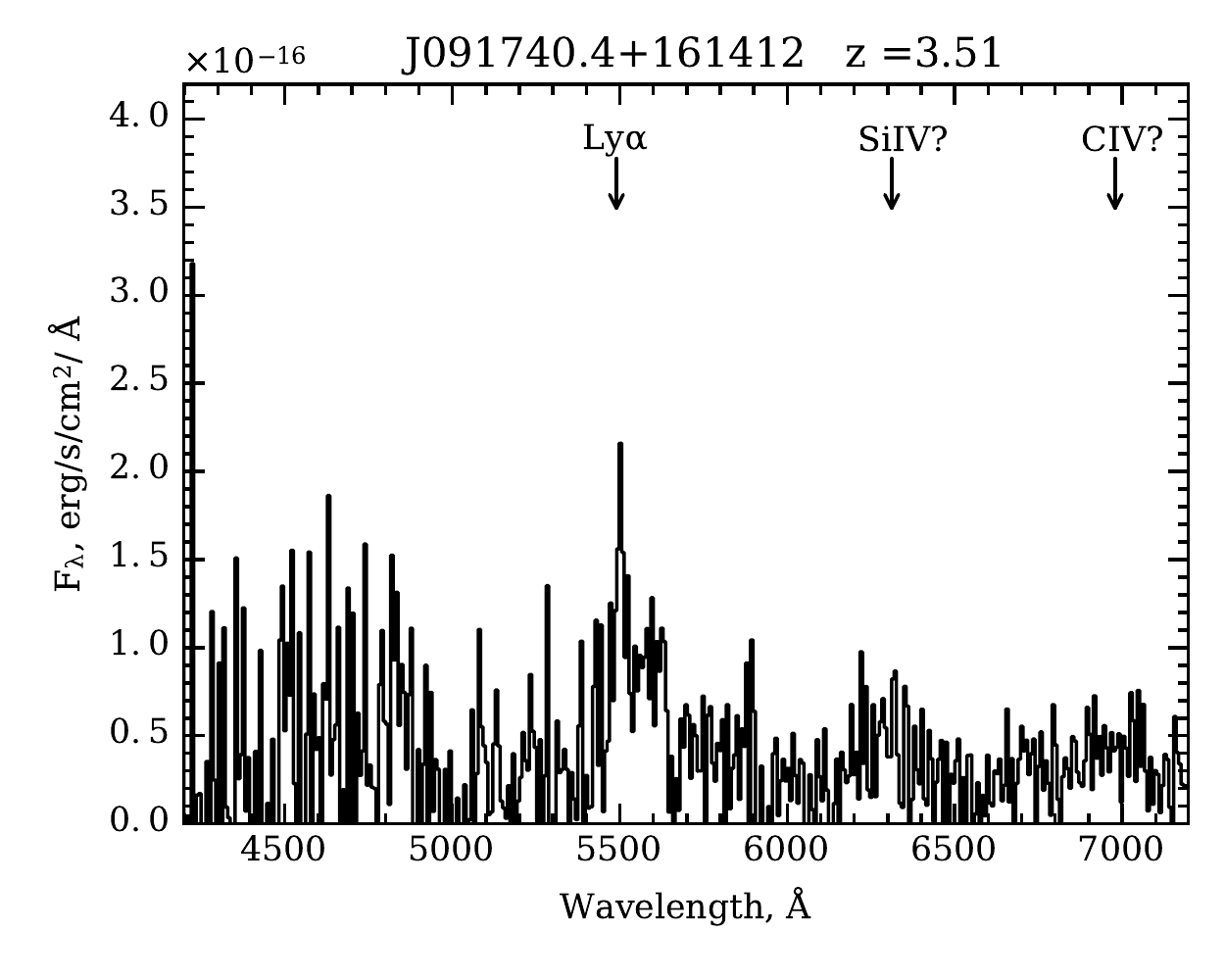}
\includegraphics[width=0.45\linewidth]{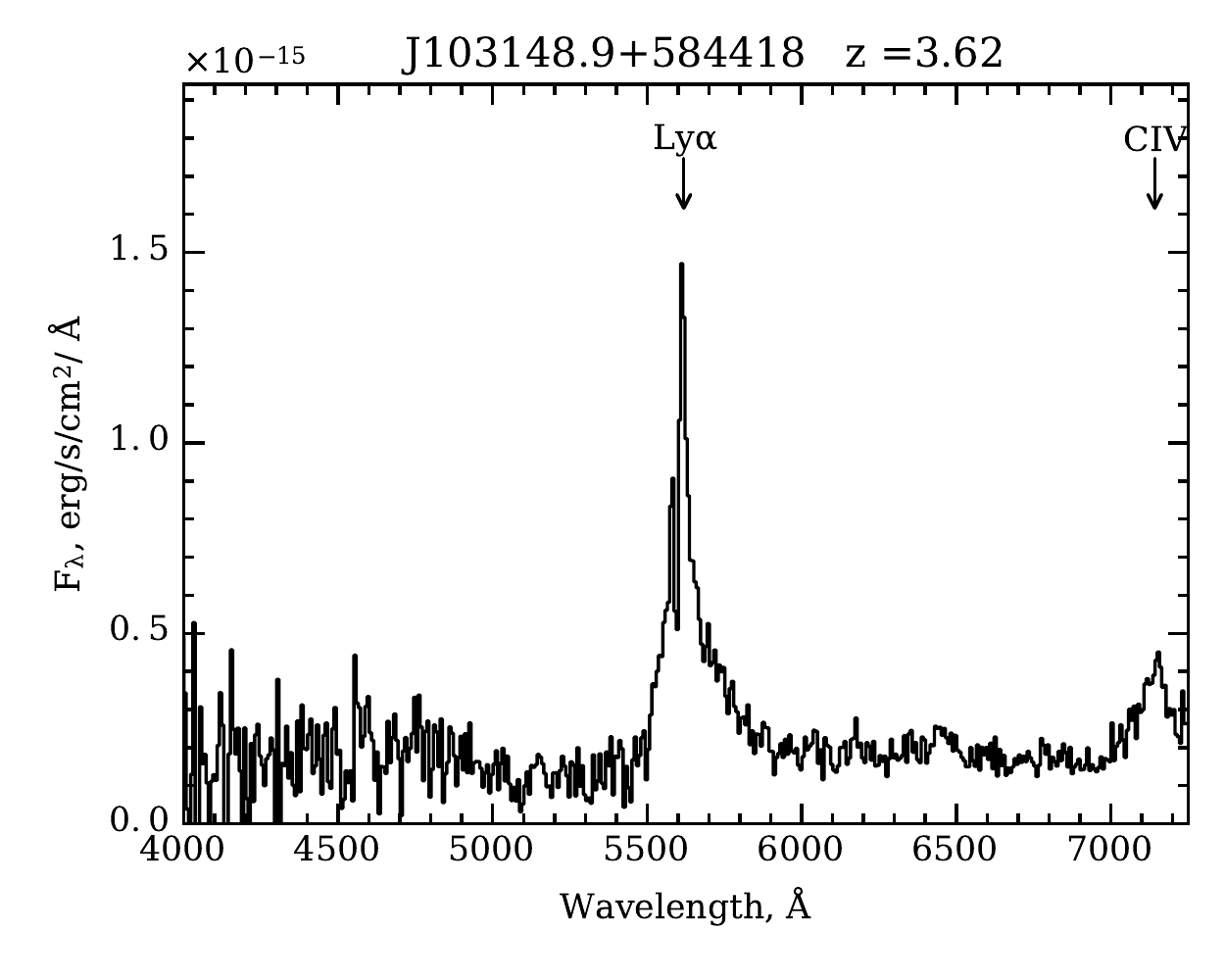}
\includegraphics[width=0.45\linewidth]{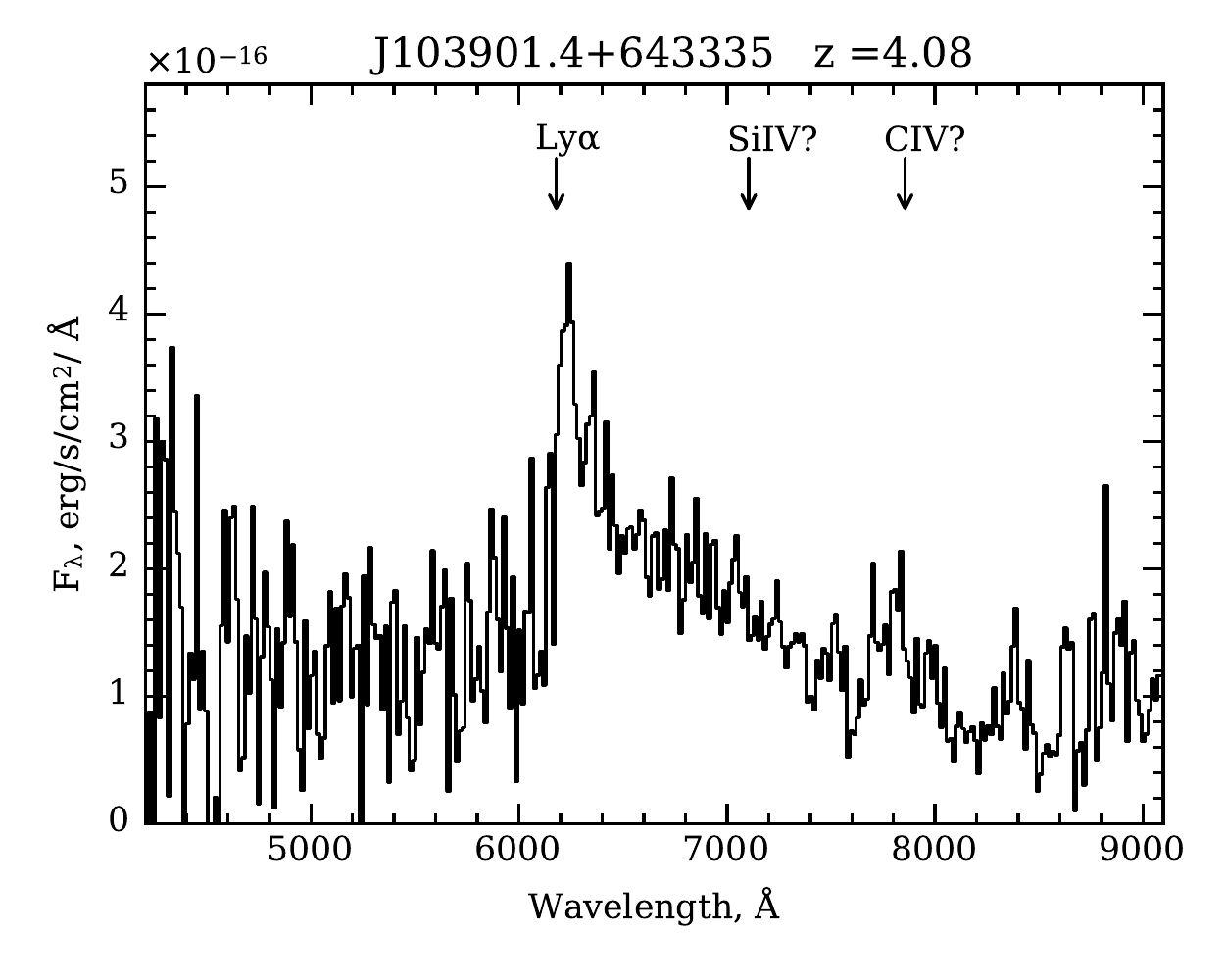}
\includegraphics[width=0.45\linewidth]{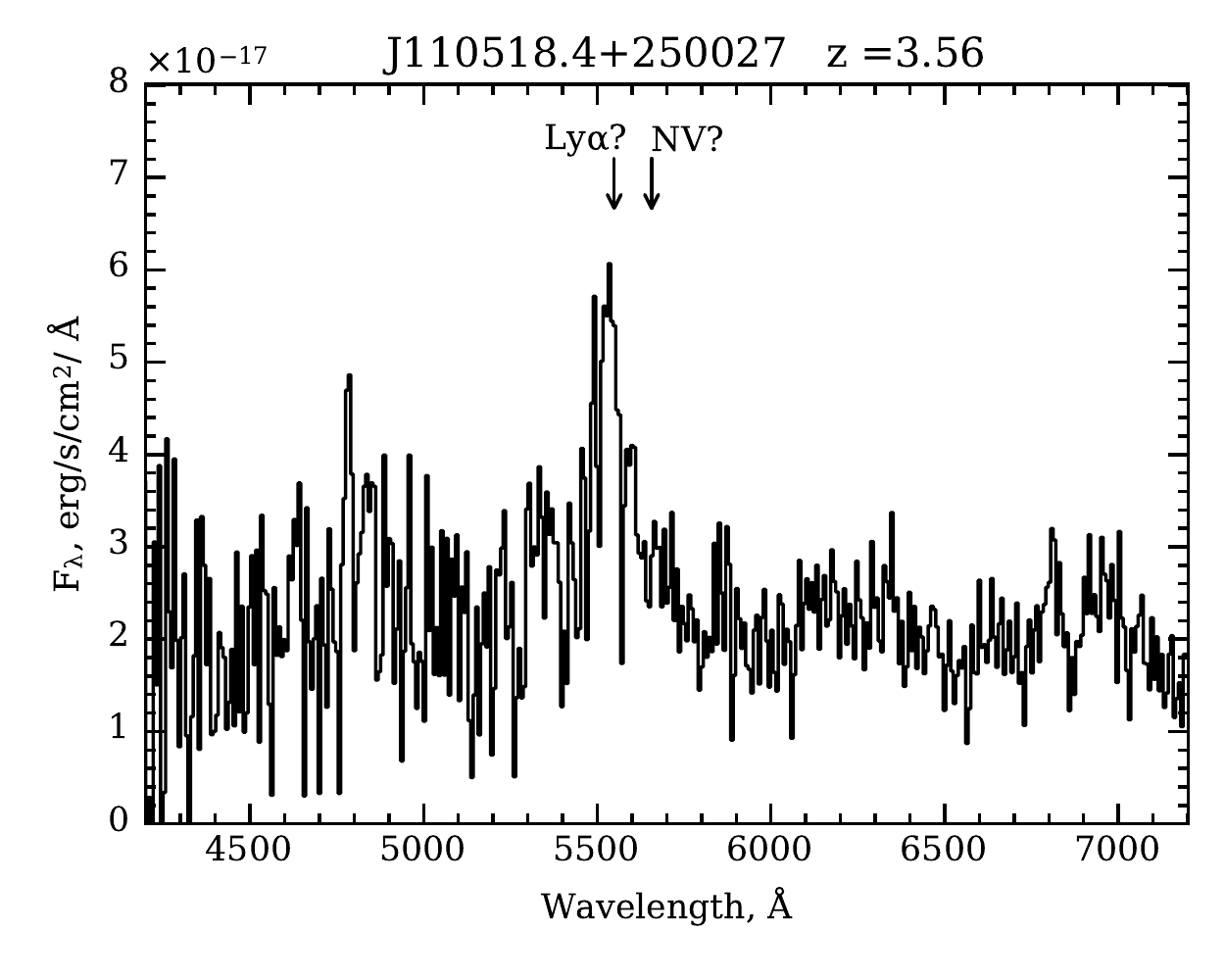}
\includegraphics[width=0.45\linewidth]{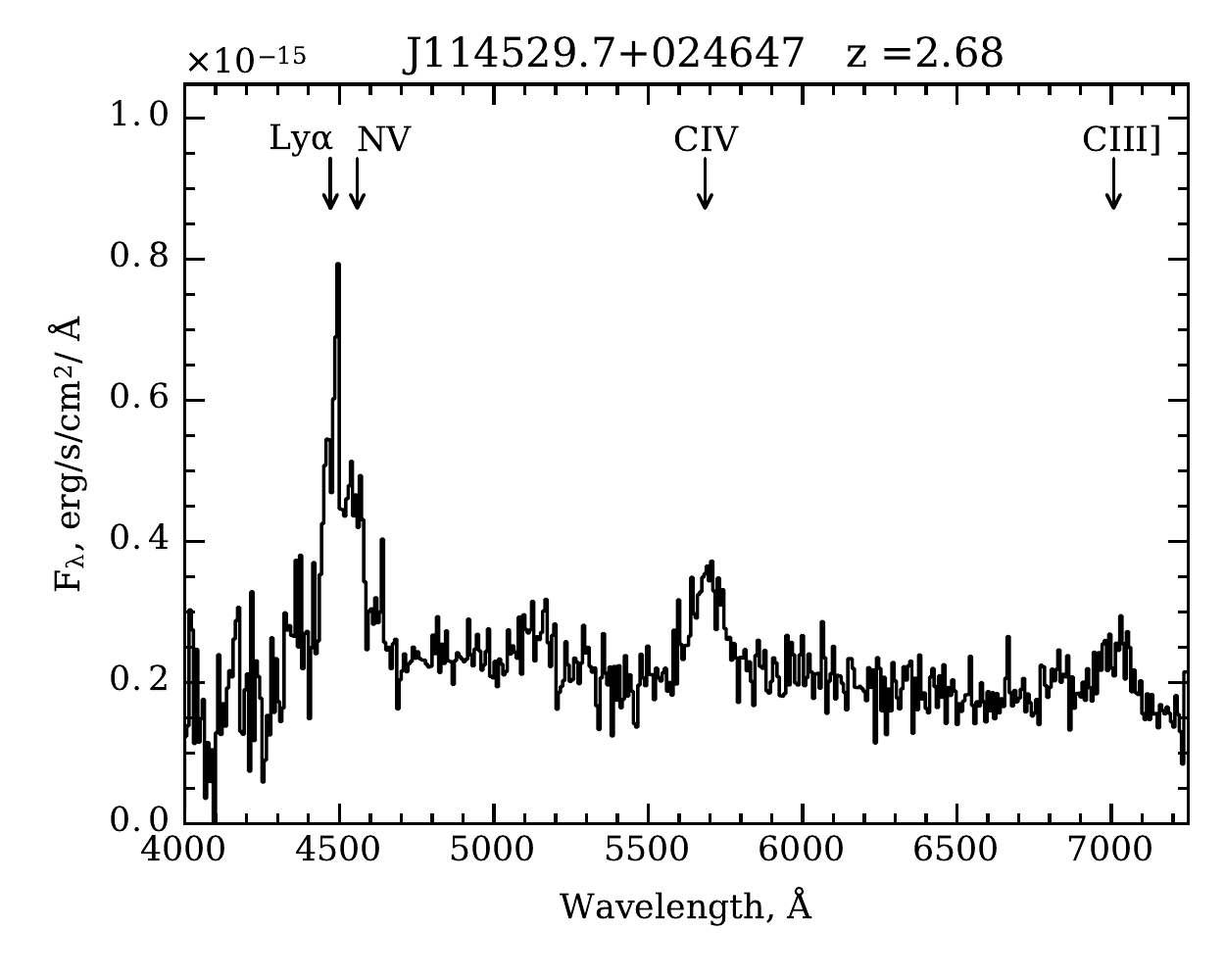}
\includegraphics[width=0.45\linewidth]{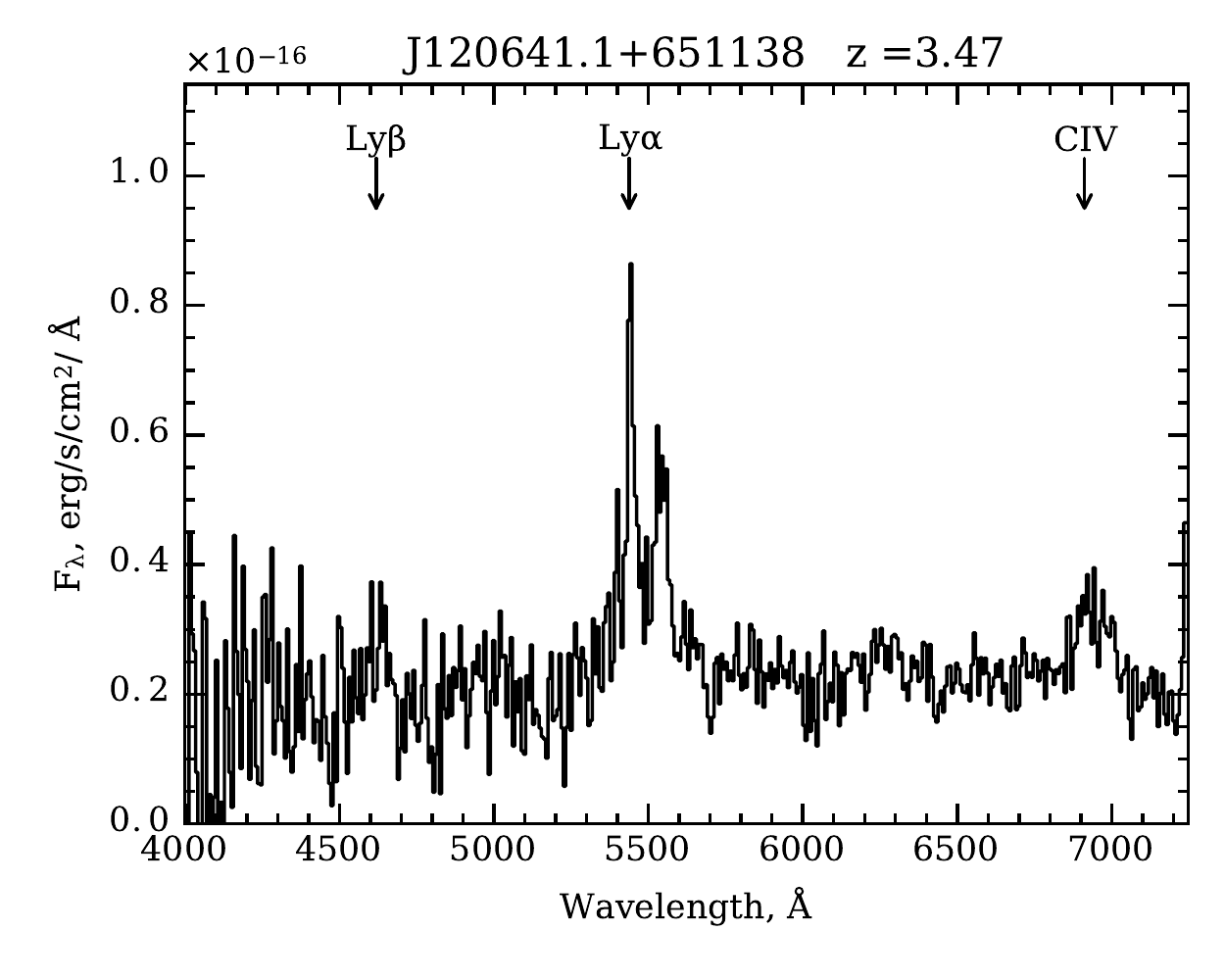}
\includegraphics[width=0.45\linewidth]{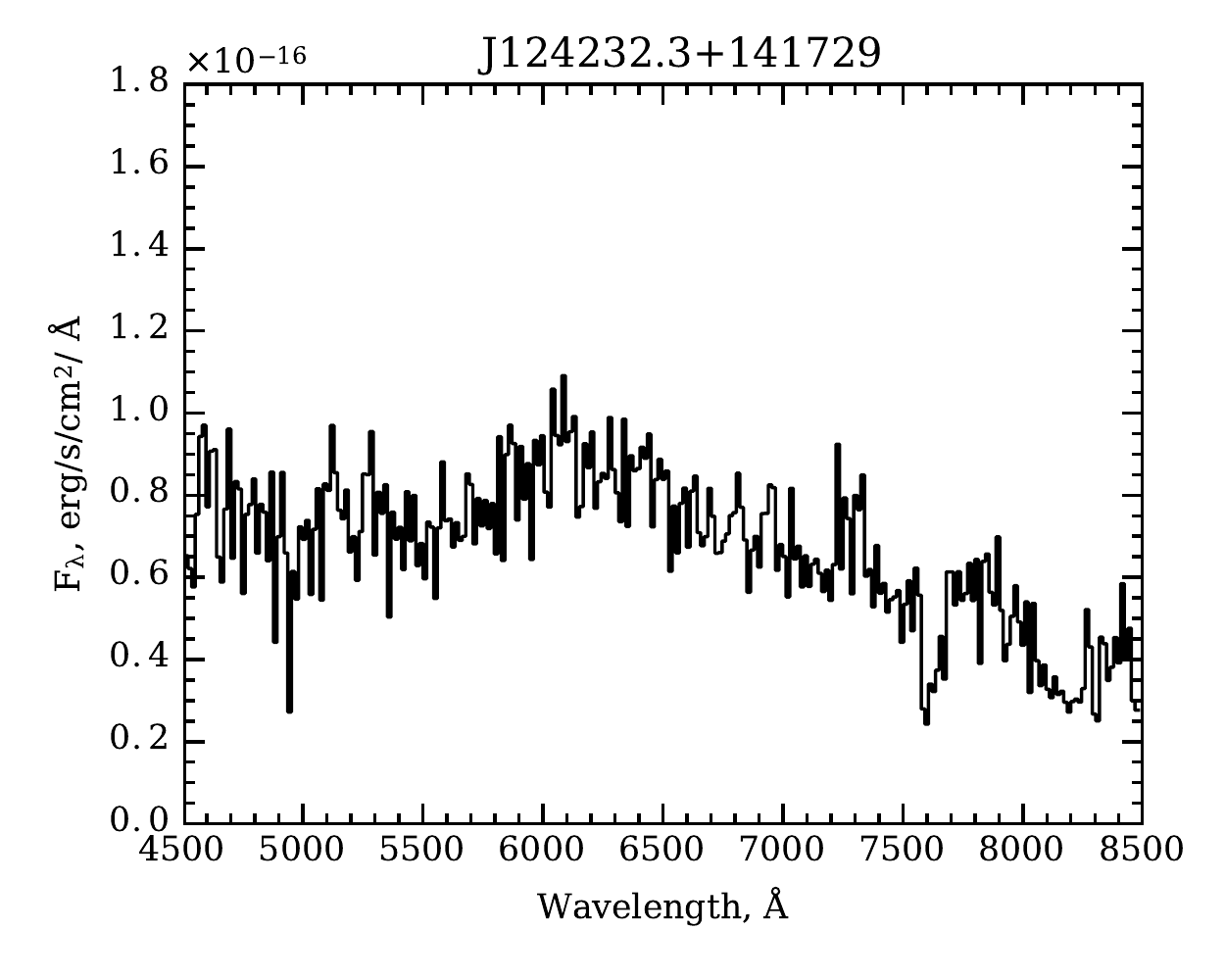}
\includegraphics[width=0.45\linewidth]{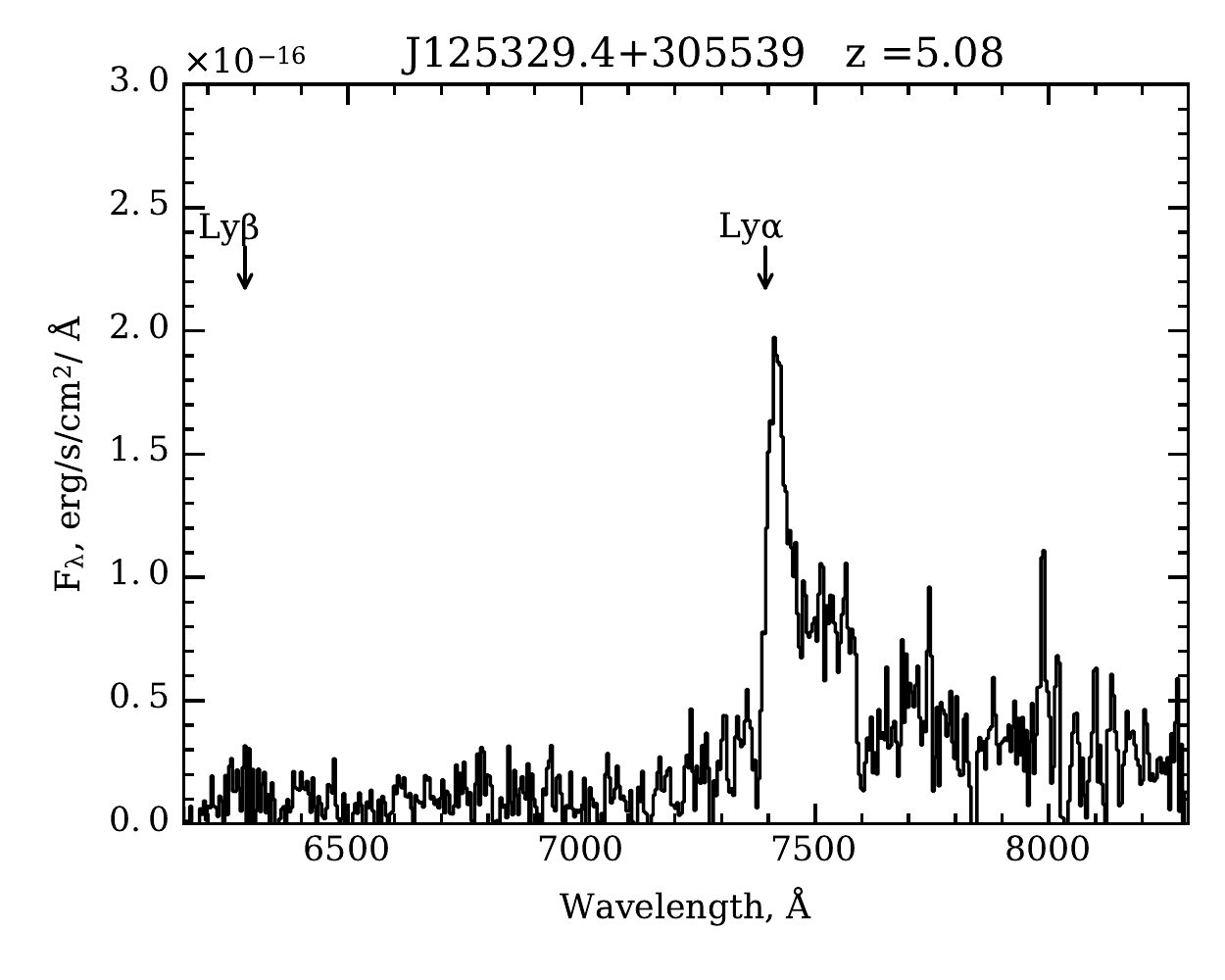}
\includegraphics[width=0.45\linewidth]{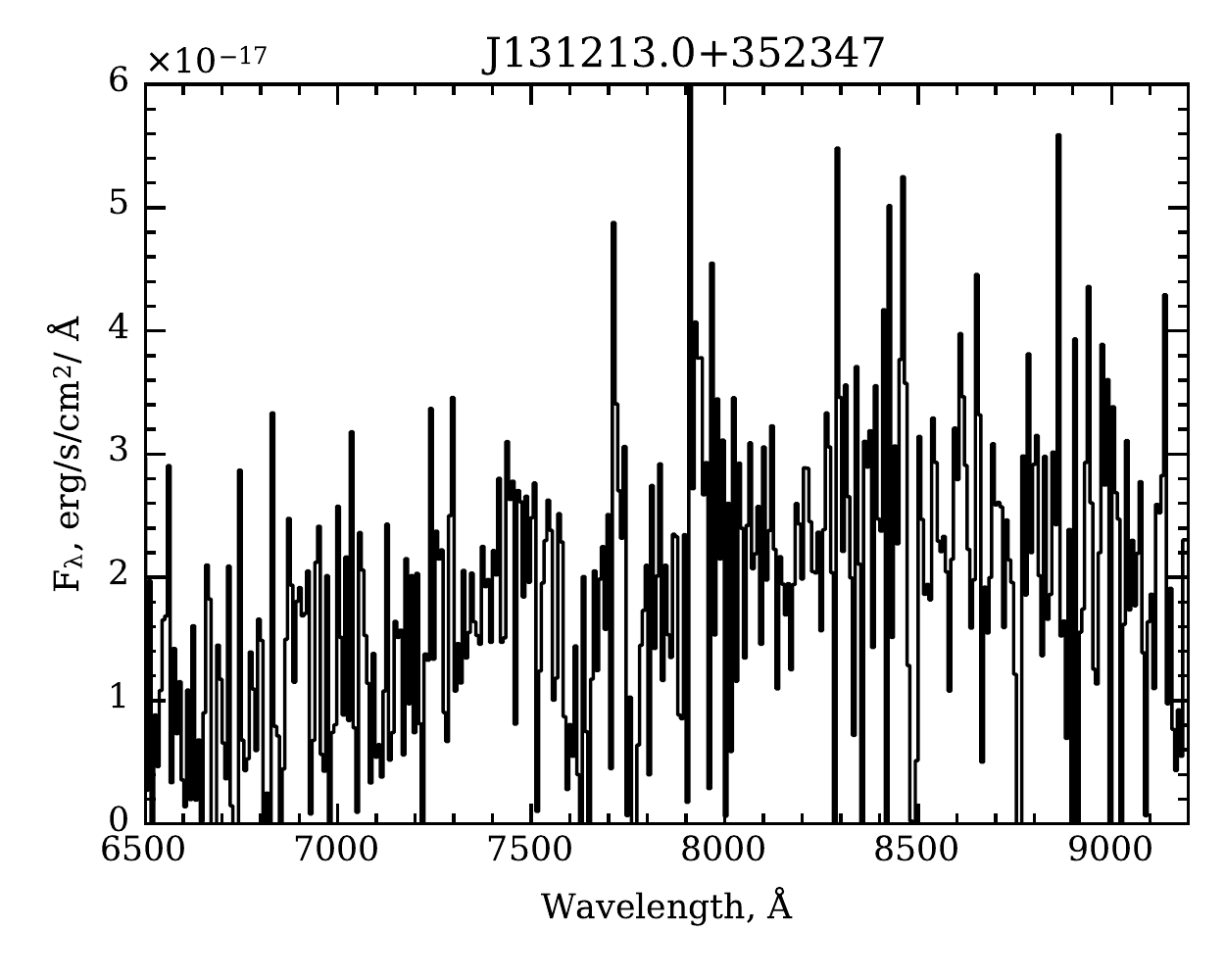}
\includegraphics[width=0.45\linewidth]{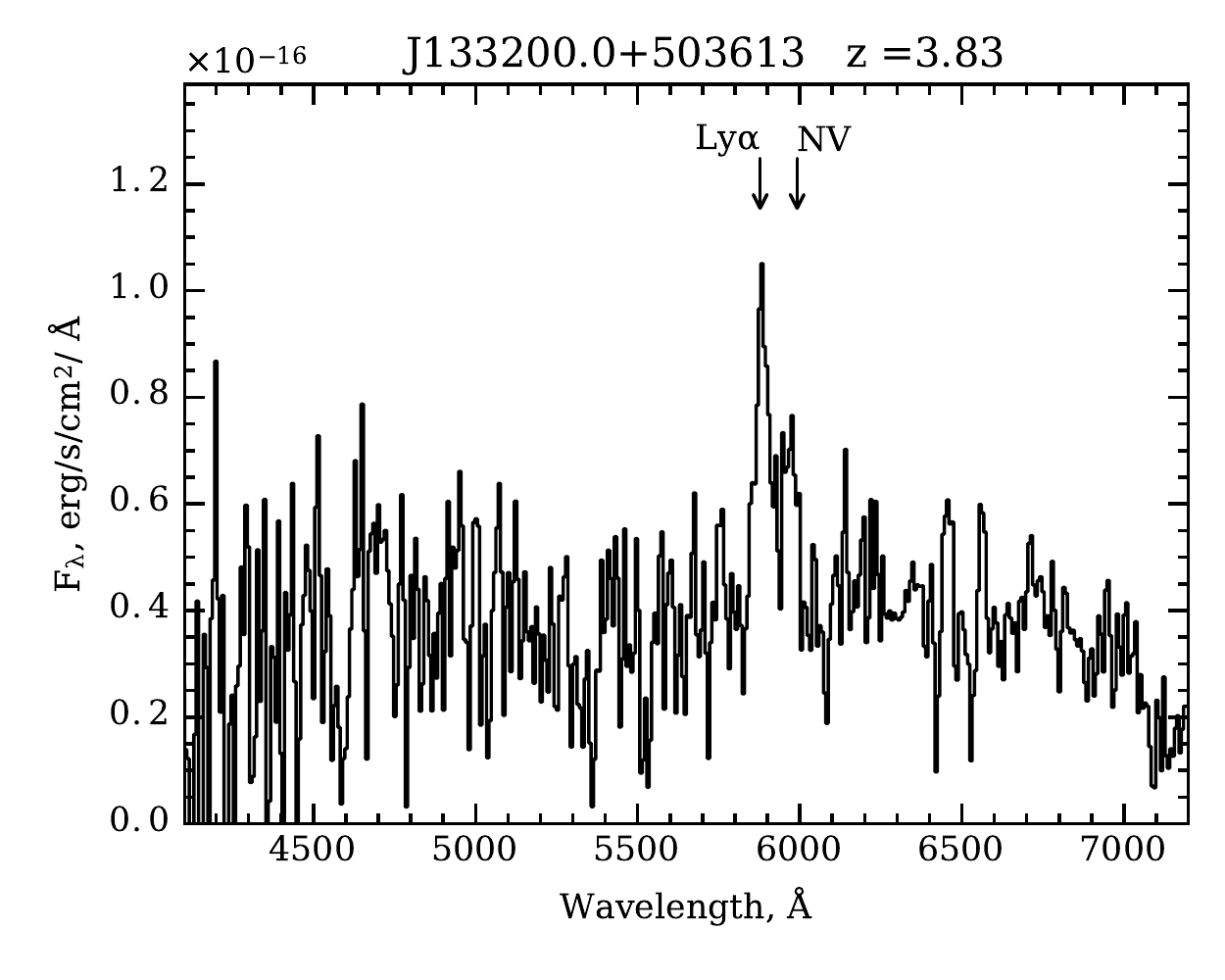}
\includegraphics[width=0.45\linewidth]{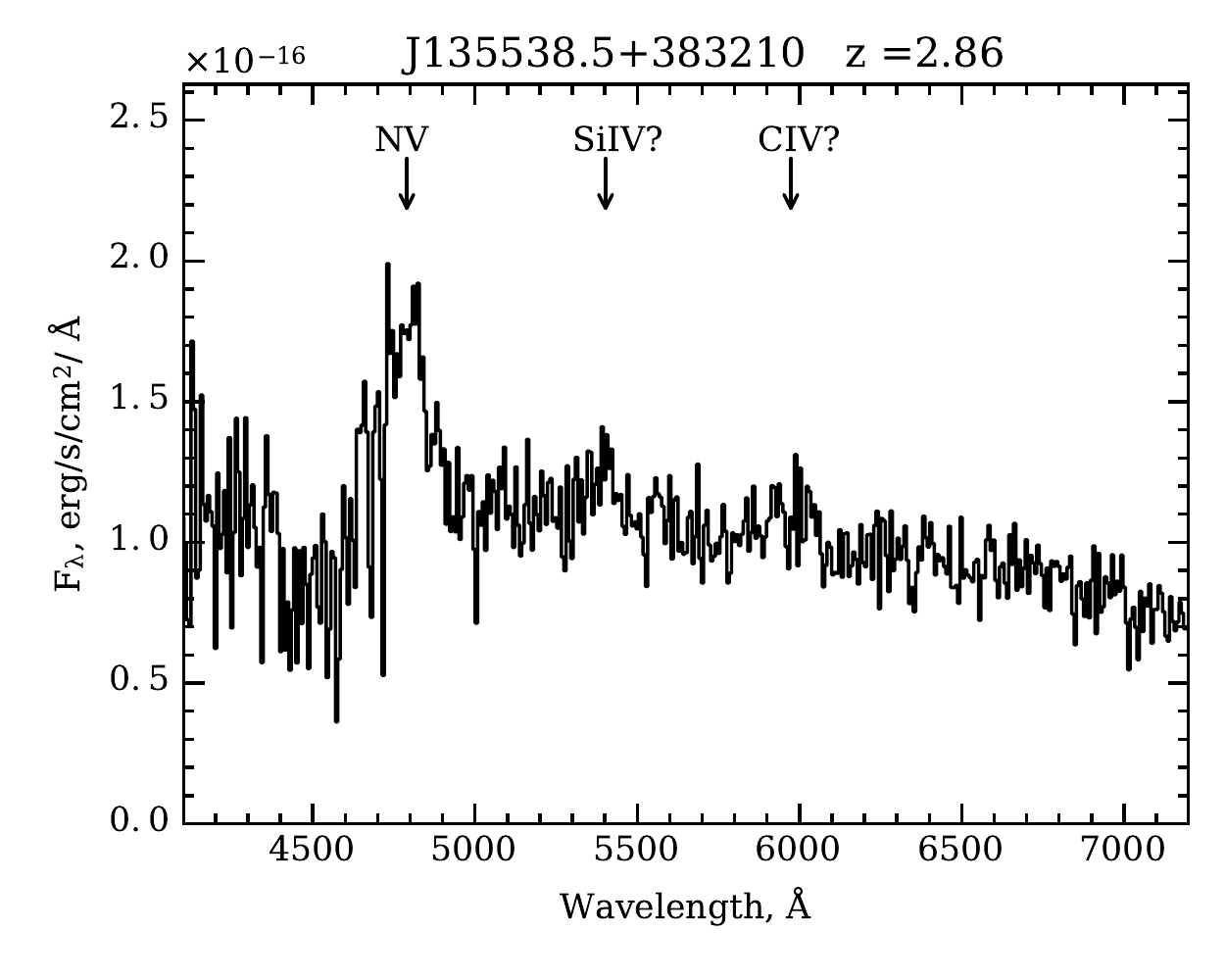}
\includegraphics[width=0.45\linewidth]{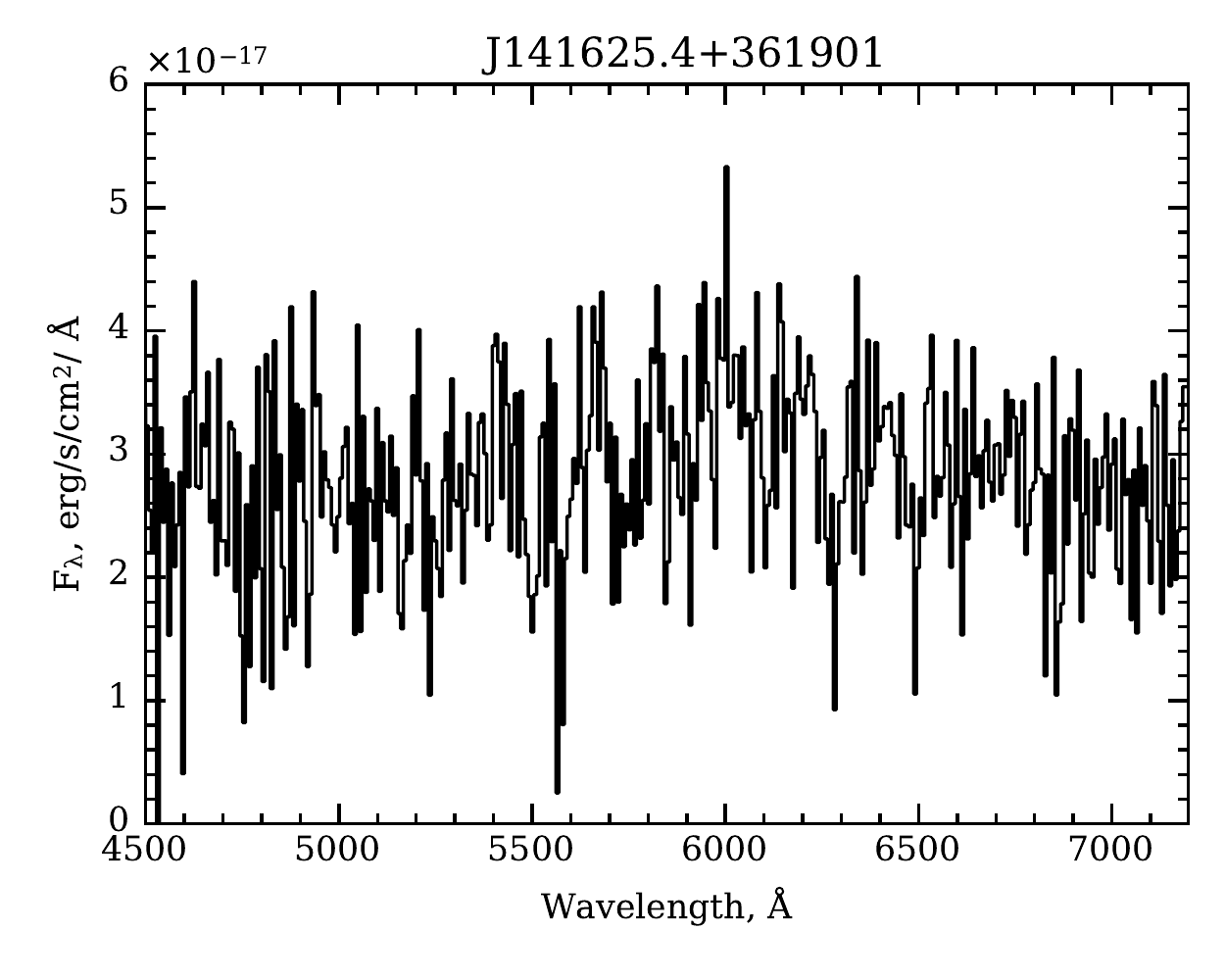}
\includegraphics[width=0.45\linewidth]{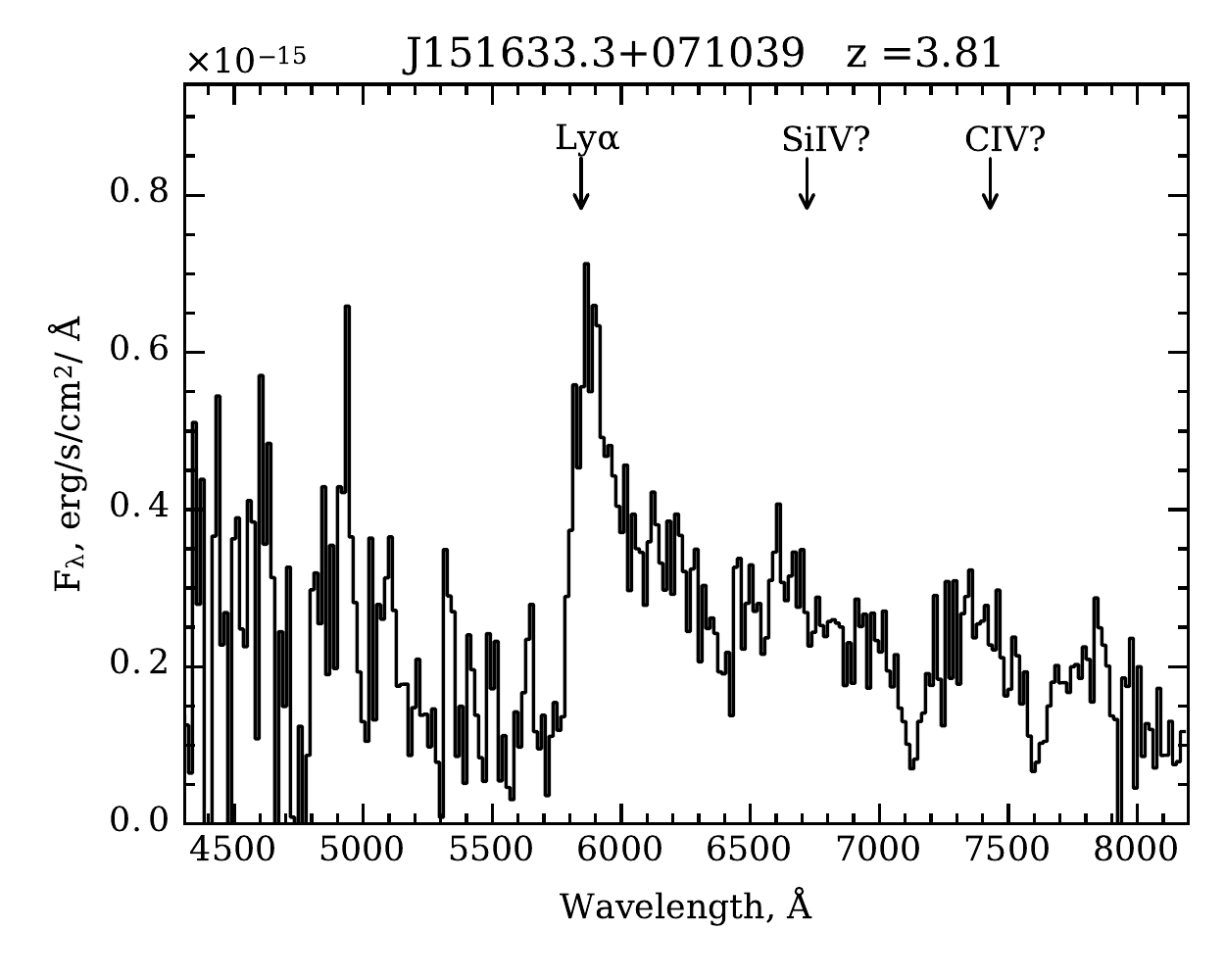}
\includegraphics[width=0.45\linewidth]{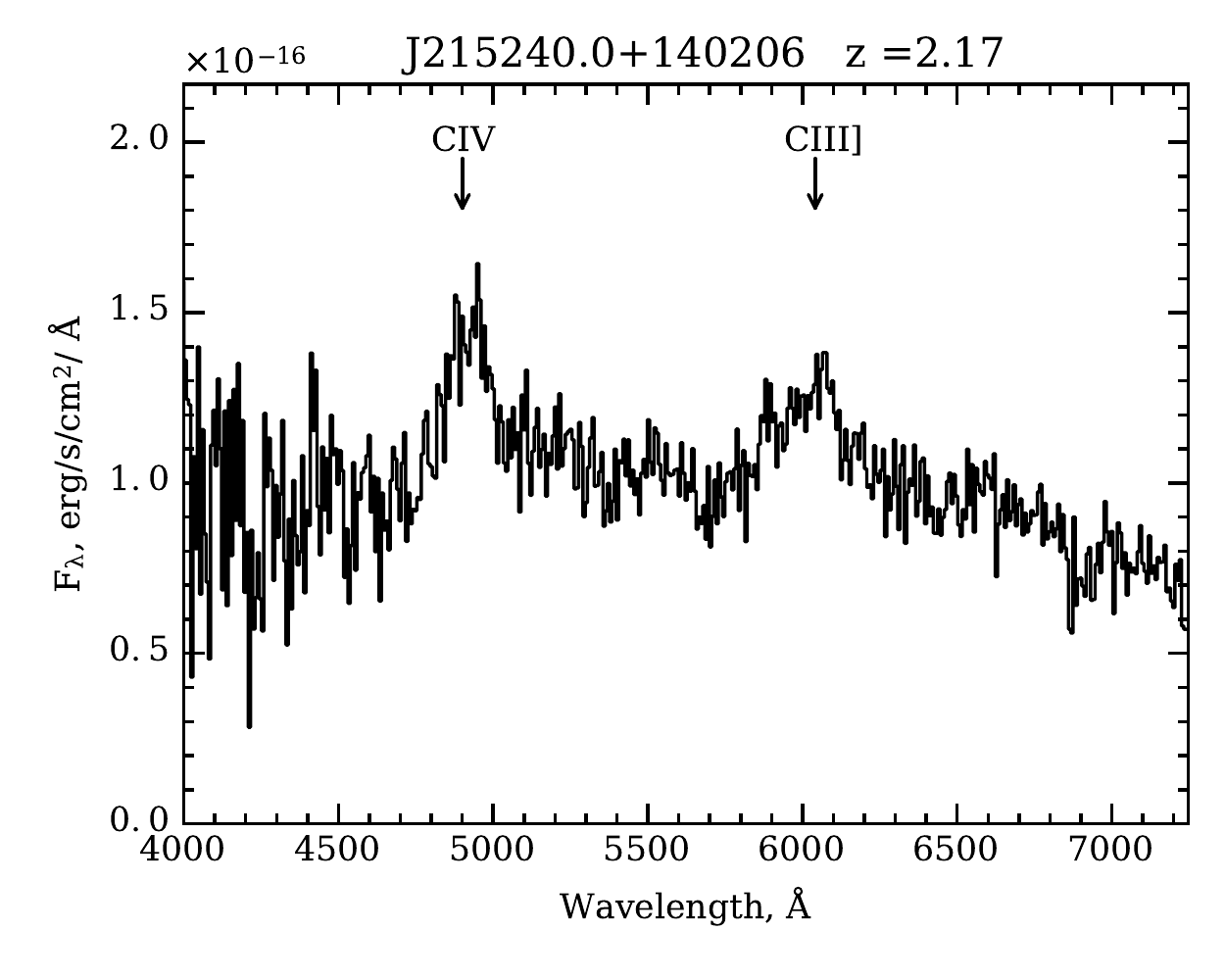}
\includegraphics[width=0.45\linewidth]{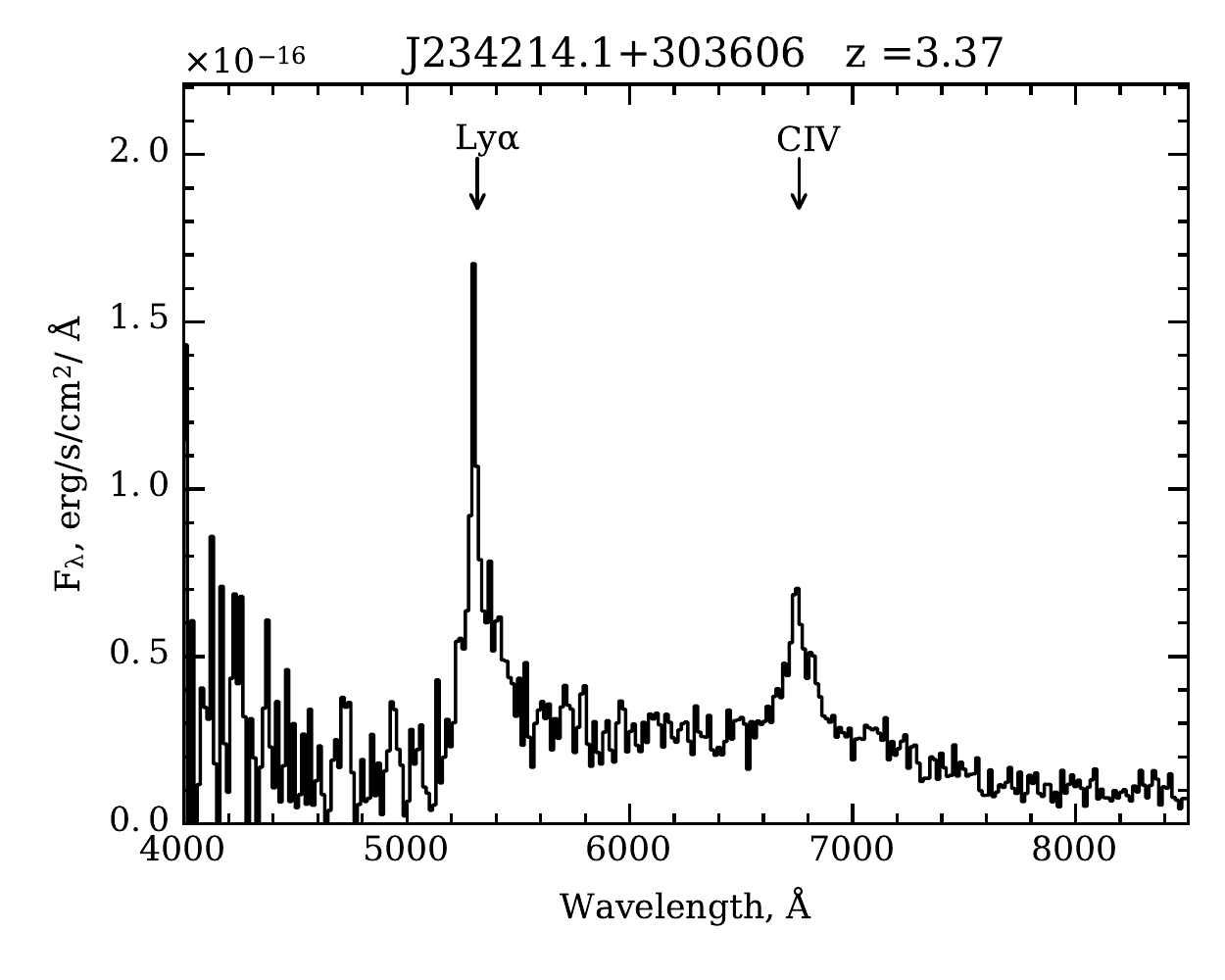}
\end{center} 
\thispagestyle{empty}

\end{document}